\documentclass[aps,amsfonts,amsmath,prd,preprint,nofootinbib,tightenlines]{revtex4-2}

\pdfoutput=1
\newcommand{\beq}{\begin{equation}}
\newcommand{\eeq}{\end{equation}}
\usepackage{graphicx}

\usepackage{xcolor}

\usepackage{tabularx}
\usepackage{multirow}
\usepackage{booktabs}
\usepackage{float}

\usepackage{amsmath}

\DeclareMathOperator\erf{Erf}

\newcommand{\be}{\begin{equation}}
\newcommand{\ee}{\end{equation}}
\newcommand{\bea}{\begin{eqnarray}}
\newcommand{\eea}{\end{eqnarray}}
\newcommand{\bi}{\begin{itemize}}
\newcommand{\ei}{\end{itemize}}

\usepackage{overpic,mathtools}

\usepackage{url}
\usepackage[colorlinks]{hyperref}
\hypersetup{ 
	colorlinks=true, 
	linkcolor=black, 
	filecolor=black, 
	citecolor = blue,       
	urlcolor=blue, 
}

\begin{document}

\title{Comparing Minimal and Non-Minimal \\Quintessence Models to 2025 DESI Data}

\author{Husam Adam$^1$\footnote{Husam.Adam@tufts.edu}, 
Mark P. Hertzberg$^{1,2}$\footnote{mark.hertzberg@tufts.edu}, 
Daniel Jim\'enez-Aguilar$^1$\footnote{Daniel.Jimenez\_Aguilar@tufts.edu}, 
Iman Khan$^{1,3}$\footnote{khani@beloit.edu}}

\affiliation{$^1$Institute of Cosmology, Department of Physics and Astronomy, Tufts University, Medford, MA 02155, USA}
\affiliation{$^2$Institute of Advanced Study, Hong Kong University of Science and Technology, Hong Kong}
\affiliation{$^3$Beloit College, Beloit, WI 53511, USA\\\\}

\begin{abstract}
In this work we examine the  2025 DESI analysis of dark energy, which suggests that dark energy is evolving in time with an increasing equation of state $w$. We explore a wide range of quintessence models, described by a potential function $V(\varphi)$, including: quadratic potentials, quartic hilltops, double wells, cosine functions, Gaussians, inverse powers. We find that while some provide improvement in fitting to the data, compared to a cosmological constant, the improvement is only modest. We then consider non-minimally coupled scalars which can help fit the data by providing an effective equation of state that temporarily obeys $w<-1$ and then relaxes to $w>-1$. Since the scalar is very light, this leads to a fifth force
and to time evolution in the effective gravitational strength, which are both tightly constrained by tests of gravity.
For a very narrow range of carefully selected non-minimal couplings we are able to evade these bounds, but not for generic values.
\end{abstract}

\maketitle

\newpage

\tableofcontents

\newpage

\section{Introduction}

The accelerated expansion of the universe \cite{SupernovaSearchTeam:1998fmf,SupernovaCosmologyProject:1998vns} is attributed to the presence of dark energy, which is characterized by an equation of state $w_{\rm DE}=p_{\rm DE}/\rho_{\rm DE}< -1/3$ and constitutes, approximately, $69\%$ of the total energy density in the universe. Despite the fact that the magnitude-redshift data for type IA supernovae and the analysis of the temperature fluctuations in the Cosmic Microwave Background (CMB) strongly suggest the existence of dark energy, a fundamental description of it is still lacking. The simplest approach is to describe it as a cosmological constant, often referred to as $\Lambda$. In this case, the dark energy density remains constant in time throughout the history of the universe, with equation of state $w_{\Lambda}=-1$. Although this model (dubbed $\Lambda$CDM) has been successful in explaining the observational data until now, it requires an energy density of $\rho_{\Lambda}\sim 10^{-47}\,\text{GeV}^{4}$, which is in large tension with the expected vacuum energy in the framework of Quantum Field Theory. This discrepancy (which is 122 orders of magnitude if compared to the Planck density), is known as the cosmological constant problem \cite{RevModPhys.61.1}, which we will not attempt to solve in this work.

The $\Lambda$CDM paradigm is 
arguably in tension with
recent results from the Dark Energy Spectroscopic Instrument (DESI) collaboration, which has reported evidence for a time-evolving dark energy density \cite{DESI:2024mwx,DESI:2025zgx}. The survey combines Baryon Acoustic Oscillation (BAO) and CMB measurements, as well as three different sets of type Ia Supernovae (Sne Ia) data: Pantheon+, Union3 and DESY5. The analysis performed for each of them in the latest release indicates $2.8\sigma$, $3.8\sigma$ and $4.2\sigma$ tensions with the $\Lambda$CDM model, respectively. The DESI collaboration described the dynamical dark energy equation of state with the so-called Chevallier-Polarski-Linder (CPL) parametrization \cite{Chevallier:2000qy,Linder:2002et}:
\begin{equation}
w_{\rm DE}(a)=w_{0}+(1-a)w_{a}\,,
\label{eq:CPL}
\end{equation}
where $a$ is the scale factor and the parameters $w_{0}$ and $w_{a}$ denote the equation of state and its derivative today (a=1). Note that the cosmological constant is obtained for $w_{0}=-1$ and $w_{a}=0$. While the CPL parametrization is by no means fundamental or generic, it might be a good approximation in the range of redshifts\footnote{Redshift is related to the scale factor via $z=-1+1/a$.} to which DESI is sensitive ($0.295\leq z\leq 2.33$). Other alternatives and extensions of the CPL parametrization have also been discussed in the literature (see, for instance, \cite{Dutta:2008qn,Chiba:2009sj,Bhattacharya:2024hep,Bhattacharya:2024kxp,Nesseris:2025lke,Dinda:2025iaq,Ozulker:2025ehg,Gialamas:2025pwv}).

There is a wide variety of models that can describe dynamical dark energy. The simplest ones are canonical quintessence models, where the role of dark energy is played by a scalar field $\varphi$ (with standard kinetic structure and minimal coupling to gravity) that evolves under the influence of some potential $V(\varphi)$. However, one can anticipate that this class of theories may not be the best candidates to provide an excellent fit to the DESI data. According to DESI’s results, the central values of the CPL parameters are $w_{0, \rm central}\approx -0.7$ and $w_{a, \rm central}\approx -1$, and so $w_{\text{DE,central}}(a)\approx -1.7+a$. Therefore, if we were to extrapolate the CPL parametrization all the way to the past, we would find $w_{\rm DE}<-1$ for $a<0.7$, or in other words, for $z>0.43$. This means that the Null Energy Condition (NEC) would be violated for such redshifts, but this cannot happen for minimally coupled scalar fields with canonical kinetic terms.\footnote{The equation of state parameter can be written as $w_{\rm DE}=-1+(p_{\varphi}+\rho_{\varphi})/\rho_{\varphi}$. The NEC states that $p_{\varphi}+\rho_{\varphi}>0$, and thus $w_{\rm DE}>-1$. For a minimally coupled scalar field, this is always true as $p_{\varphi}+\rho_{\varphi}=\dot{\varphi}^{2}$. However, we can relax this for non-minimally coupled scalars, as we discuss later.} Since $z=0.43$ lies inside DESI’s range of redshifts, one concludes that an accurate fit to the DESI data is not possible, although there could still be a modest level of compatibility.\footnote{In models that consider an extended parameter space, this level of compatibility can be higher (see, for instance, \cite{RoyChoudhury:2024wri,RoyChoudhury:2025dhe}).} 
Values of $w_{0}$ and $w_{a}$ closer to the central ones may be obtained in theories where NEC violation is not forbidden (or is just apparent). These include models of scalar fields non-minimally coupled to gravity \cite{Ye:2024ywg,Wolf:2024stt,Wolf:2025jed,Wang:2025znm} (see also \cite{Giacomini:2020grc,Andriot:2025los}), braneworld dark energy \cite{Mishra:2025goj}, dark matter-dark energy interactions \cite{Li:2024qso,Li:2025owk,Chakraborty:2025syu,vanderWesthuizen:2025iam,Guedezounme:2025wav,Samanta:2025oqz} (see \cite{Wang:2024vmw} for a detailed review), fields with non-standard kinetic terms \cite{Goldstein:2025epp,Chen:2025ywv} or modified gravity theories \cite{Dimakis:2023cam,Yang:2025mws,Paliathanasis:2025hjw,Paliathanasis:2025xxm,Nojiri:2025low,Hogas:2025ahb}. Other possibilities are based on non-standard equations of state for dark matter \cite{Kumar:2025etf,Chen:2025wwn,Braglia:2025gdo} and modified cosmologies in the framework of holographic dark energy \cite{Luciano:2025hjn} or the generalized uncertainty principle \cite{Paliathanasis:2025dcr}. To date, the strongest evidence for a phantom crossing in the equation of state of dark energy has been reported in \cite{Scherer:2025esj} with an approximate significance level of $5\sigma$.

In this paper, we will extend the work presented in \cite{Bayat:2025xfr} for other quintessence potentials, and we will also consider scalar field models with a non-minimal coupling to gravity. Our goal is to assess whether the DESI data significantly favors these theories over the $\Lambda$CDM model.

The manuscript is structured as follows. We study a set of canonical quintessence models characterized by different scalar potentials in Sec. \ref{sec:models}, and provide their statistical analysis in Sec. \ref{sec:statistical analysis}. Then, in Sec. \ref{sec:nonminimal coupling}, we report on our findings for scalar fields non-minimally coupled to gravity. Finally, we present our conclusions in Sec. \ref{sec:conclusions}.

We work in natural units ($c=\hbar=1$) and use the $(-+++)$ convention for the spacetime metric. Furthermore, $M_{p}$ will denote the reduced Planck mass: $M_{p}=1/\sqrt{8\pi G}$.

\section{Models}
\label{sec:models}

We start by considering canonical quintessence models in which a single scalar field $\varphi$ plays the role of the dark energy. These models are described by the following action:
\begin{equation}
S=\int d^{4}x\sqrt{-g}\left[\frac{M_{p}^2}{2}R-\frac{1}{2}g^{\mu\nu}\partial_{\mu}\varphi\partial_{\nu}\varphi-V(\varphi)+\mathcal{L}_{\rm m}\right]\,,
\label{eq:action}
\end{equation}
where $R$ is the Ricci scalar, $V(\varphi)$ is the potential energy density of the scalar field and $\mathcal{L}_{\rm m}$ denotes the Lagrangian density of the matter fields. 

Focussing on an FLRW universe, the equation of motion for the scalar field reads
\begin{equation}
\ddot{\varphi}+3H\dot{\varphi}+V'(\varphi)=0\,.
\label{eq:eom}
\end{equation}
Moreover, the evolution of the scale factor is governed by the Friedmann equation:
\begin{equation}
H^{2}=\frac{1}{3M_{p}^{2}}\left(\rho_{\rm m}+\rho_{\varphi}\right)\,,
\label{eq:Friedmann}
\end{equation}
where $H=\dot{a}/a$ is the Hubble rate and $\rho_{\rm m}$ and $\rho_{\varphi}$ respectively denote the energy densities of matter (proportional to $a^{-3}$) and the scalar field. The latter is given by
\begin{equation}
\rho_{\varphi}=\frac{\dot{\varphi}^{2}}{2}+V(\varphi)\,.
\label{eq:energy density phi}
\end{equation}
Since we are mostly interested in late-time dynamics, we are neglecting the contribution of radiation. Note also that we are considering a spatially flat universe.

In the following subsections, we numerically solve (\ref{eq:eom}) and (\ref{eq:Friedmann}) for different scalar field potentials. Regarding the initial conditions, the field will be released from rest\footnote{We will assume that  the initial Hubble rate is high enough to keep the field frozen at early times. Later when we study non-minimally coupled models, this is more subtle.} ($\dot{\varphi}_{i}=0$) at different values $\varphi_{i}$. This completely determines the initial energy density of the scalar field. On the other hand, the initial energy density of matter will be chosen to be much greater than that of the scalar field.


\subsection{Potentials}

\begin{figure}[t]
\centering
\includegraphics[scale=0.63]{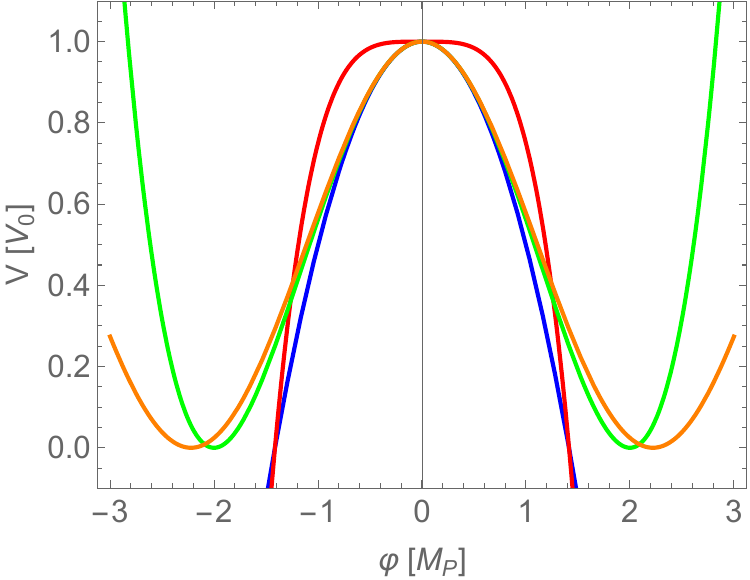}
\,\,
\includegraphics[scale=0.63]{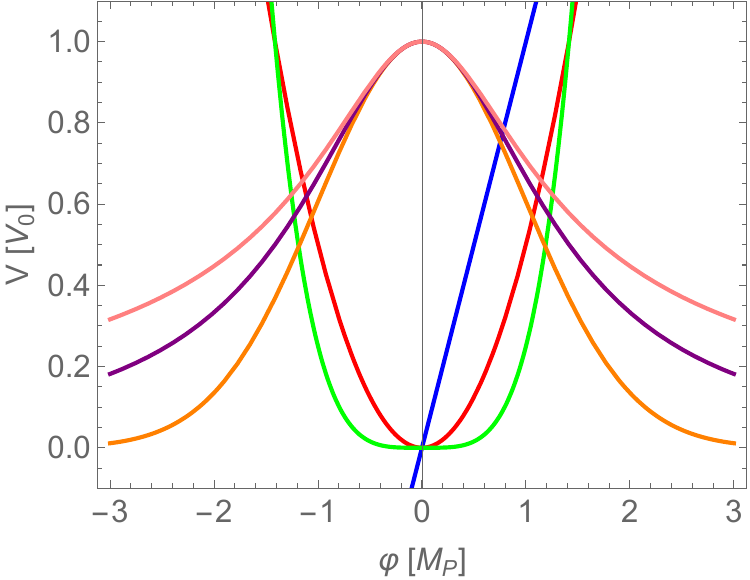}
\caption{Potentials $V(\varphi)$ considered in this work.
Left: Blue is quadratic hilltop eq.~(\ref{V1}), 
red is quartic hilltop eq.~(\ref{V2}),
green is double well with $\lambda=1$ eq.~(\ref{V3}), 
orange is cosine eq.~(\ref{V4}). 
Right: 
Blue is linear monomial eq.~(\ref{V10}), 
red is quadratic monomial eq.~(\ref{V5}), 
green is quartic monomial eq.~(\ref{V6}),
orange is Gaussian eq.~(\ref{V7}), 
purple is inverse function eq.~(\ref{V8}),
pink is inverse square root function eq.~(\ref{V9}). 
We have set $k=1$ throughout.}
\label{fig:potentials}
\end{figure}

The quintessence models are specified by a potential $V(\varphi)$.
Since $\varphi$ is a Lorentz scalar, so is any function $V(\varphi)$. Hence there are infinitely many options for the potential.
To make progress, we will consider a range of potential functions.
Those considered in this work are shown in Fig. \ref{fig:potentials}.

\begin{figure}[t]
\centering
\includegraphics[scale=0.48]{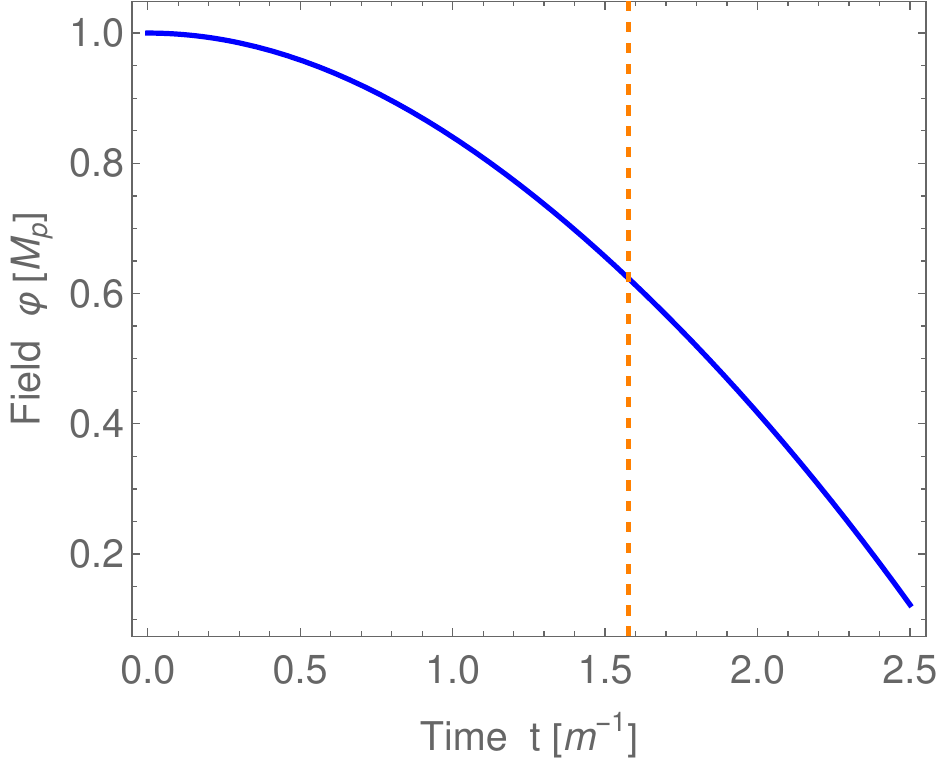}\,\,
\includegraphics[scale=0.48]{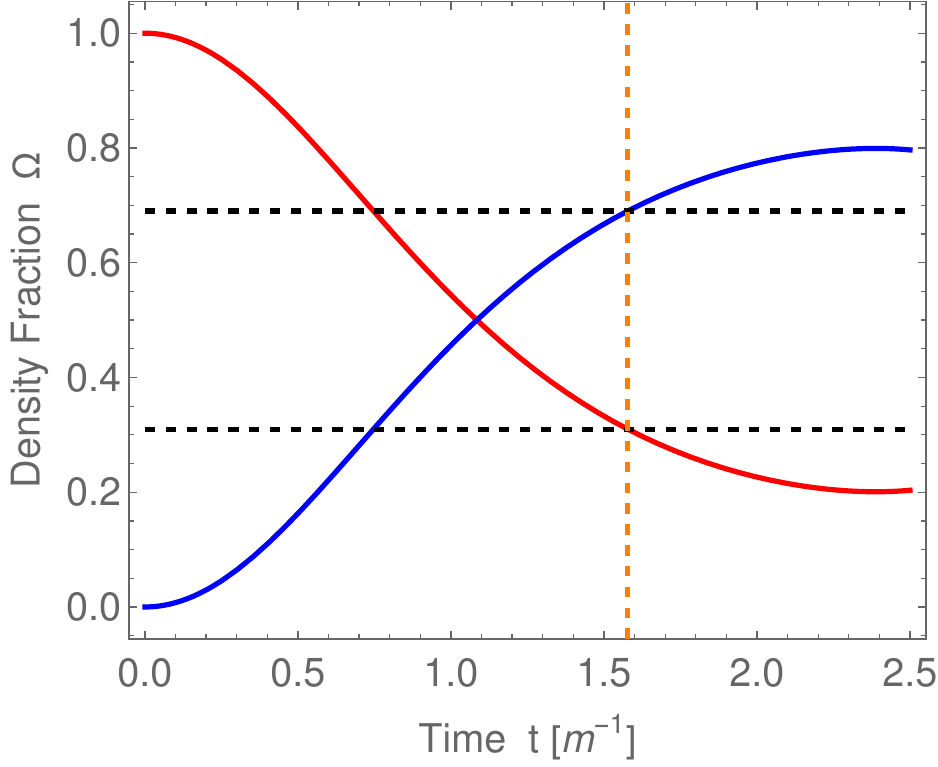}\\
\vspace{0.5cm}
\includegraphics[scale=0.54]{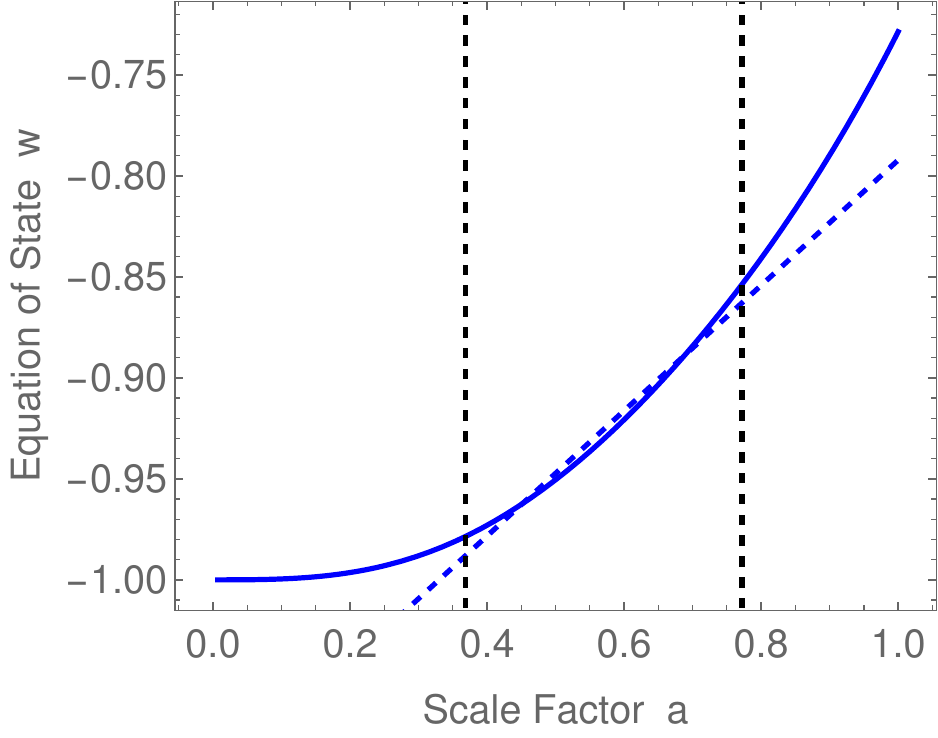}
\caption{
Top left: Value of the scalar field as a function of time. 
Top right: Fractional energy densities of matter (red) and dark energy (blue).
The vertical dashed line (orange) corresponds to the current time, defined as the moment at which the dark energy density fraction becomes $\Omega_{DE,0}=0.69$, which in this case is $t_0\approx 1.58 m^{-1}$. 
Bottom: Dark energy equation of state $w$ as a function of the scale factor (solid curve) and CPL fit (blue dashed line). The vertical lines indicate the range over which DESI is most sensitive. The present time is at $a=1$.
These plots are for the linear potential eq.~(\ref{V10})
with $\varphi_i=M_p$.}
\label{fig:simple}
\end{figure}

For each value of $\varphi_i$ and parameters in the potential we evolve the system. We then output the equation of state $w$, and take its best fit value between redshifts $0.295\leq z\leq 2.33$. The procedure is indicated in Fig.~\ref{fig:simple} for the upcoming case of the linear potential.

\subsubsection{Hilltop Potentials}

In Ref.~\cite{Bayat:2025xfr} there was a focus on models near a hilltop, in which the potential is a constant minus a quadratic. We will reexamine this here (eq.~(\ref{V1})) in light of the updated 2025 DESI data.
Moreover, we will also consider a range of other hilltop potentials:
a quartic hilltop (eq.~(\ref{V2})), as well as models with local minima (eqs.~(\ref{V3},\,\ref{V4})). For the double well in eq.~(\ref{V3}), it has a late-time vacuum energy that vanishes if $\lambda=1$, but is non-zero otherwise. We note that since these models all have dynamical dark energy, we will not need to run them to very late times; so it is not directly important if the (very) late time vacuum energy is positive, negative, or vanishing.
The set of potentials considered is:
\begin{eqnarray}
V(\varphi)&=&V_{0}\left(1-\frac{k^{2}}{2M_{p}^2}\varphi^{2}\right)
\label{V1}\\
V(\varphi)&=&V_{0}\left(1-\frac{k^{4}}{4M_{p}^4}\varphi^{4}\right)
\label{V2}\\
V(\varphi)&=&V_{0}\left(1-\frac{k^{2}}{2M_{p}^2}\varphi^{2} +\frac{\lambda k^4}{16M_{p}^4}\varphi^{4}\right)
\label{V3}\\
V(\varphi)&=&\frac{V_{0}}{2}\left(1+\cos\left(\frac{\sqrt{2}\,k\,\varphi}{M_{p}}\right)\right)
\label{V4}
\end{eqnarray}
Here, the overall scale that sets the density $V_0$ will be rewritten in terms of a characteristic mass scale $m$ via $V_0=m^2M_p^2$. Going forwards, we will measure time in units of $m^{-1}$. These potentials are expressed in terms of a dimensionless parameter $k$, which controls the curvature at the hilltop. For $k\ll1$, the field rolls very slowly off the hilltop. For  $k\gg1$, the field rolls very quickly off the hilltop. In the latter case, to have a long-lived universe, one should have the field begin very close to the hilltop, so this still takes quite some time.

\begin{figure}[t]
    \centering
    \includegraphics[width=0.48\textwidth]{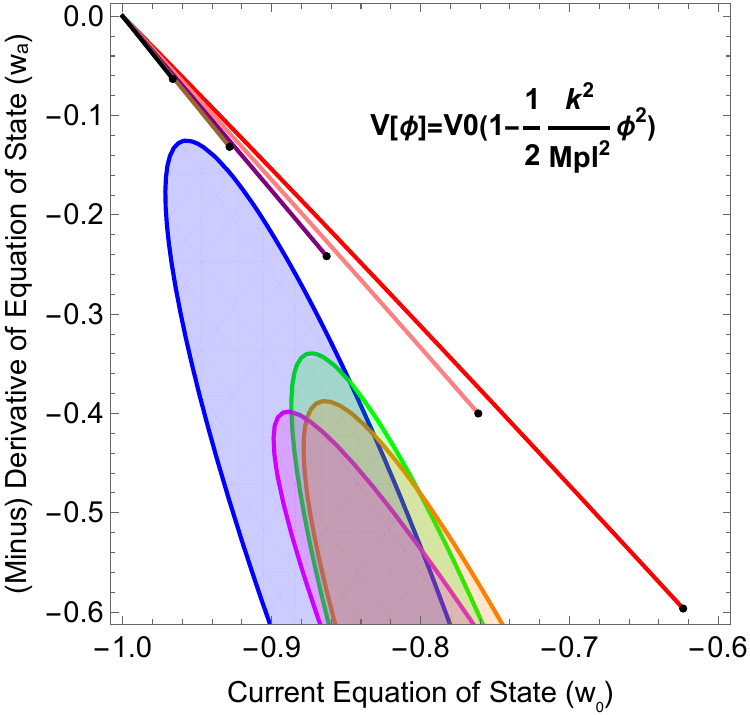}
      \includegraphics[width=0.48\textwidth]{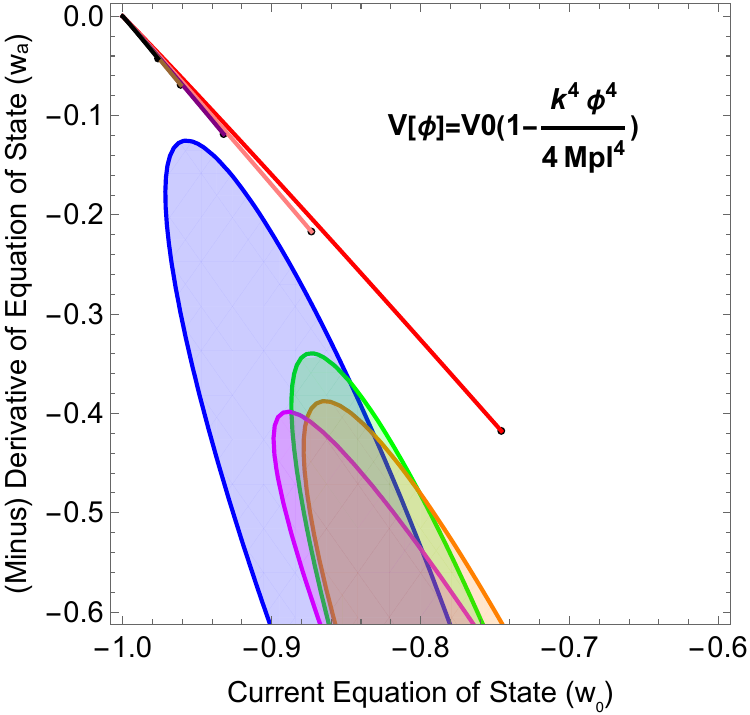}\\
      \vspace{0cm}
        \includegraphics[width=0.48\textwidth]{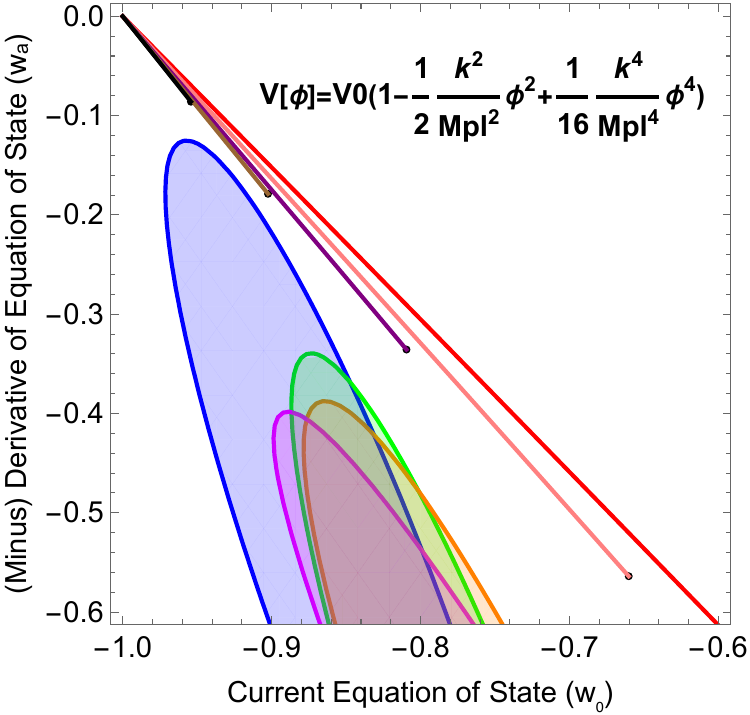}
        \includegraphics[width=0.48\textwidth]{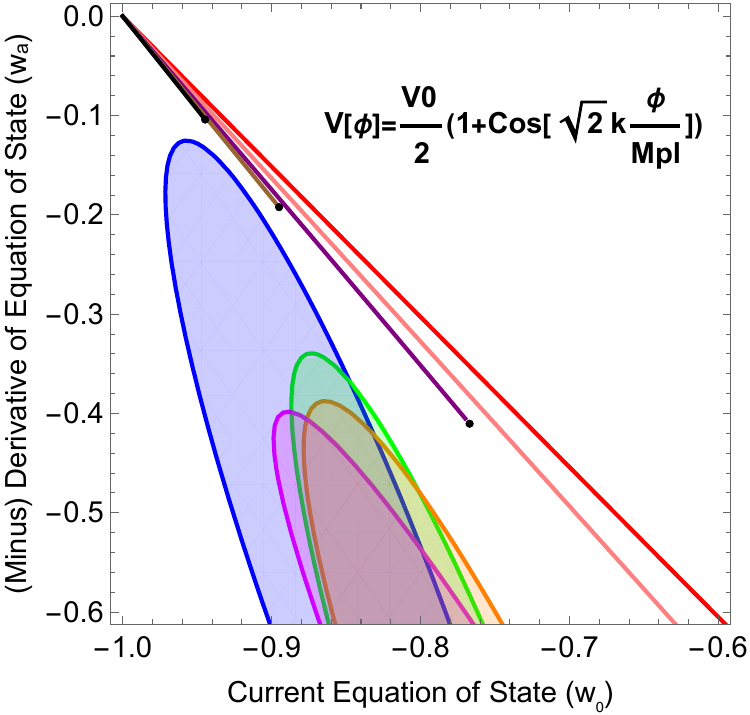}
    \caption{Equation of state parameters $w_a$ versus $w_0$. 
    Top left is quadratic hilltop potential eq.~(\ref{V1}).
    Top right is quartic hilltop potential eq.~(\ref{V2}).
    Bottom left is double well potential with $\lambda=1$ eq.~(\ref{V3}).
    Bottom right is cosine potential eq.~(\ref{V4}).
    Red is $k = 1$, pink is $k = 2$, purple is $k = 3$, brown is $k = 4$, and black is $k = 5$. 
    We are varying $\varphi_i$ up to its maximum value. 
    Also, the three different contours correspond approximately to the 95\% region of the datasets DESI BAO + CMB + PantheonPlus (blue), DESI BAO + CMB + Union3 (orange), DESI BAO + CMB + DESY5 (green), and DESI BAO + CMB (magenta) from Ref.~\cite{DESI:2025zgx}.}
    \label{fig:QuadraticHilltop}
\end{figure}

Taking the values from the best fit for $w_0,\,w_a$, we report our results for this set of models in Fig.~\ref{fig:QuadraticHilltop}. Each curve is for a fixed value of $k$, with $\varphi_i$ varied. For $\varphi_i\to0$, the field is sitting extremely close to the hilltop and so it acts as a cosmological constant ($w_0=-1,\,w_a=0$), while for larger $\varphi_i$ it becomes dynamical. There is typically a maximum value for $\varphi_i$; beyond this, the field rolls so fast that the dark energy density never grows to be $\Omega_{DE,0}=0.69$. So we end the curves at these points (this is indicated by the dots at the ends of the lines). Overall we see that all the curves do not go through the data; though some get close.

\subsubsection{Monomial Potentials}

\begin{figure}[t]
    \centering

    \includegraphics[width=0.48\textwidth]{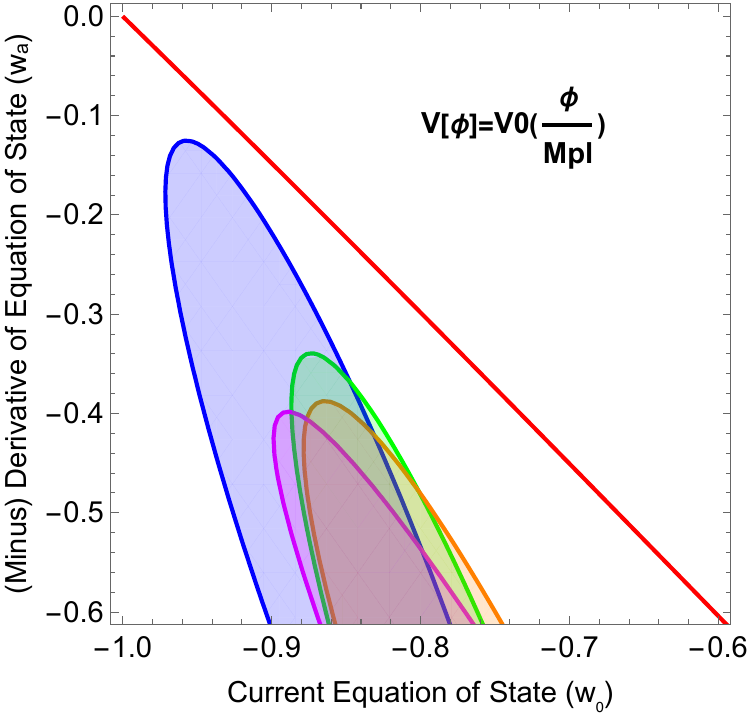}\\
    \includegraphics[width=0.48\textwidth]{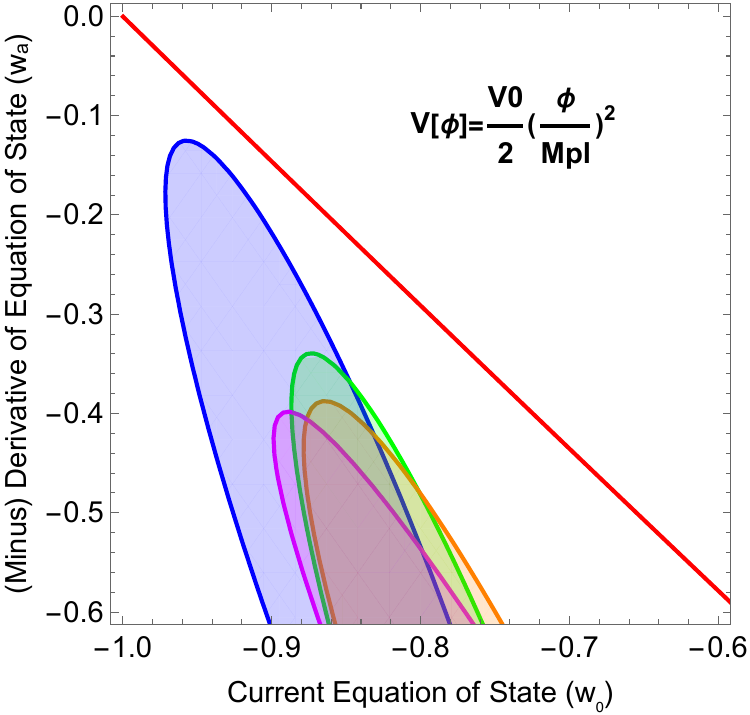}
        \includegraphics[width=0.48\textwidth]{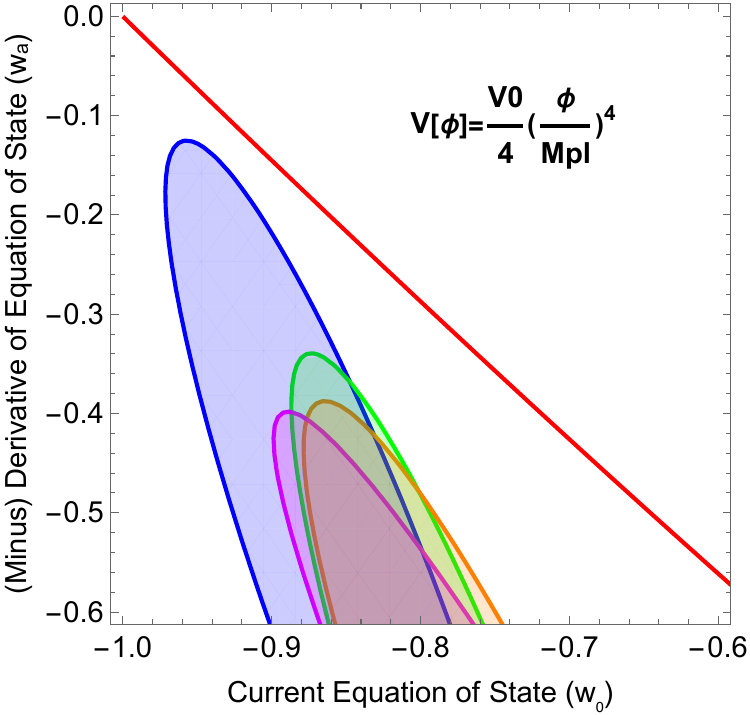}
        \caption{Equation of state parameters $w_a$ versus $w_0$. 
    Top is pure linear potential eq.~(\ref{V10}). Left is pure quadratic potential eq.~(\ref{V5}).
    Right is pure quartic potential eq.~(\ref{V6}).
    We are varying $\varphi_i$ from its minimum value. 
    Also, the three different contours correspond approximately to the 95\% region of the datasets DESI BAO + CMB + PantheonPlus (blue), DESI BAO + CMB + Union3 (orange), DESI BAO + CMB + DESY5 (green), and DESI BAO + CMB (magenta) from Ref.~\cite{DESI:2025zgx}.
    }
    \label{fig:PureQuadratic}
\end{figure}

We also consider models which are a pure (positive) power law -- a monomial. 
If the field starts near the minimum, then it just oscillates for even powers or becomes negative for odd powers, and does not act as dark energy at all. So there is a minimum value of $\varphi_i$ here. For the linear potential it is $\varphi_{\rm min,i}\approx0.85M_p$; for the 
quadratic monomial it is $\varphi_{\rm min,i}\approx 1.4M_p$; for the quartic monomial it is $\varphi_{\rm min,i}\approx 2.4 M_p$.
On the other hand, if the field starts at $\varphi_i\gg M_p$, then the evolution is similar to that of a cosmological constant. So the intermediate regime in which $\varphi_i\sim\mbox{few}\times M_p$ is of most interest for dynamical dark energy. The power laws considered in this work are:
\begin{eqnarray}
V(\varphi)&=&V_0\left(\varphi\over M_p\right)\label{V10}\\
V(\varphi)&=&\frac{V_0}{2}\left(\varphi\over M_p\right)^2\label{V5}\\
V(\varphi)&=&\frac{V_0}{4}\left(\varphi\over M_p\right)^4\label{V6}
\label{eq:monomial}
\end{eqnarray}
By measuring time in units of $m^{-1}$ (recall $V_0=m^2M_p^2$), the only remaining parameter in these models is the initial $\varphi_i$ (in units of $M_p$). The results for this analysis are given in Fig.~\ref{fig:PureQuadratic}. We note that in this case, both the pure quadratic and the pure quartic give rather similar results. And none of the results really cuts into the DESI data.

\subsubsection{Decaying Potentials}

Another family of potentials we consider are those that decay asymptotically towards $V\to0$ at large $\varphi$. For small $\varphi$, these are similar to the hilltop models considered earlier. However, for large $\varphi$, there is a new type of behavior. 
In eq.~(\ref{V7}) there is rapid decay from a Gaussian.
In eq.~(\ref{V8}) there is inverse power law decay.
In eq.~(\ref{V9}) there is inverse square root decay.
The potentials are:
\begin{eqnarray}
V(\varphi)&=&V_{0}\,\exp\!\left(-{\frac{k^2\varphi^2}{2M_{p}^2}}\right)
\label{V7}\\
V(\varphi)&=&V_0\left(1+\frac{k^2\varphi^2}{2M_{p}^2}\right)^{\!-1}
\label{V8}\\
V(\varphi)&=&V_0{\left(1+\frac{k^2\varphi^2}{M_{p}^2}\right)}^{\!-1/2}
\label{V9}
\end{eqnarray}

\begin{figure}[h!]
    \centering
     \includegraphics[width=0.48\textwidth]{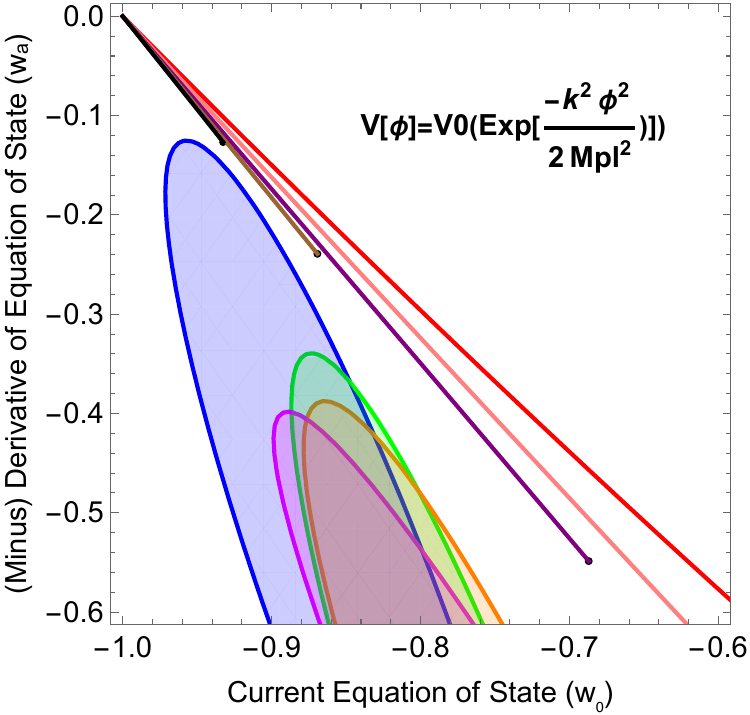}\\
     \vspace{0.4cm}
    \includegraphics[width=0.48\textwidth]{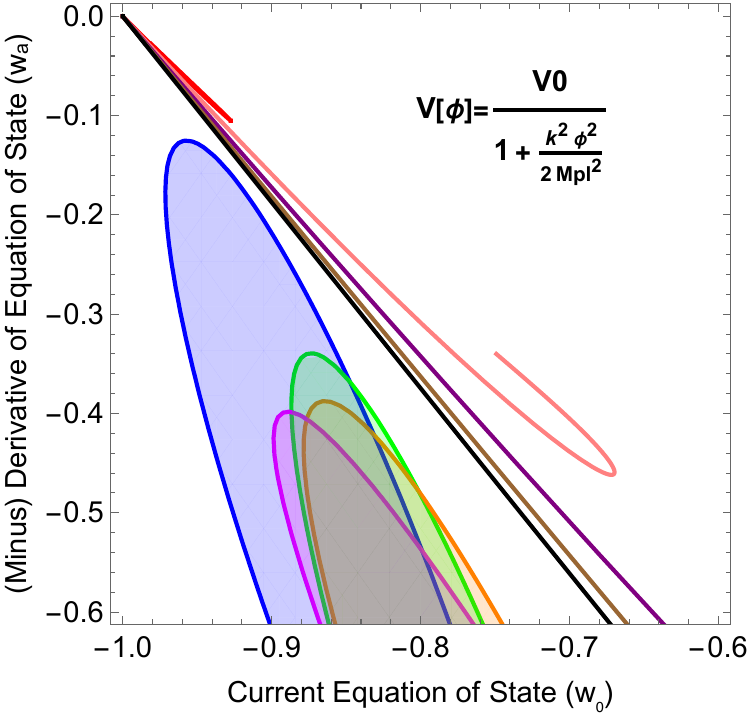}
       \includegraphics[width=0.48\textwidth]{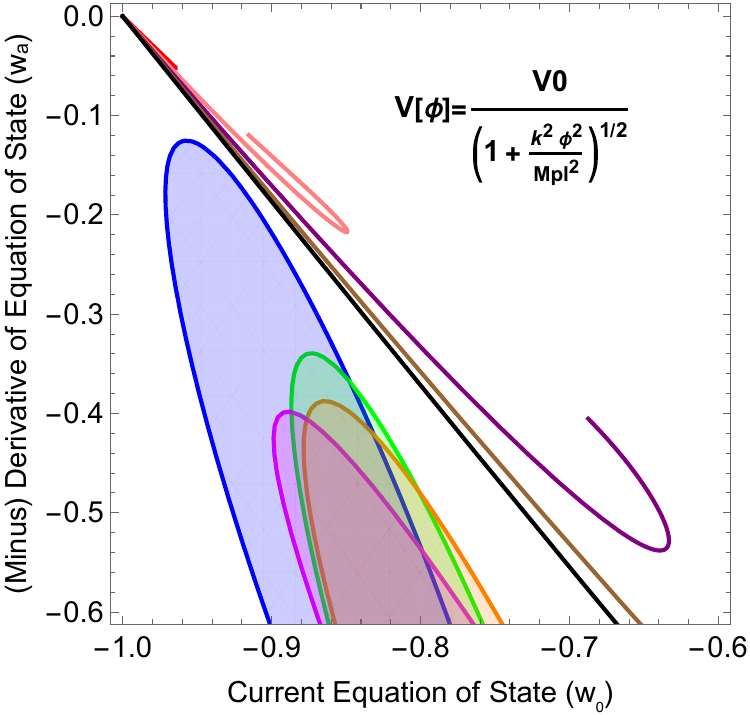}
    \caption{Equation of state parameters $w_a$ versus $w_0$. 
    Top is Gaussian potential eq.~(\ref{V7}).
    Bottom left is inverse potential eq.~(\ref{V8}).
    Bottom right is inverse square root potential eq.~(\ref{V9}).
    Red is $k = 1$, pink is $k = 2$, purple is $k = 3$, brown is $k = 4$, and black is $k = 5$. 
    We are varying $\varphi_i$ to large values. The inverse potentials continue to loop around to the cosmological constant point.
    Also, the three different contours correspond approximately to the 95\% region of the datasets DESI BAO + CMB + PantheonPlus (blue), DESI BAO + CMB + Union3 (orange), DESI BAO + CMB + DESY5 (green), and DESI BAO + CMB (magenta) from Ref.~\cite{DESI:2025zgx}.}
    \label{fig:Inverse}
\end{figure}

We note that in the power law decay cases, there is qualitatively new behavior at large $\varphi$. The results in the $w_0,\,w_a$ plane are given in Fig.~\ref{fig:Inverse}. At large $\varphi$ we see that the shape of the curves turn around. This is because there are now two limits in which we have a cosmological constant: small $\varphi_i$ and large $\varphi_i$. We see that the turn-around is in the wrong direction relative to the data, so the fit is still not accurate. But going forward, we wish to be precise.

\section{Statistical Analysis}
\label{sec:statistical analysis}

In order to assess the statistical preference for the quintessence models described above, we follow the procedure described in Ref. \cite{Bayat:2025xfr}. Firstly, the joint probability density function of $w_{0}$ and $w_{a}$ is assumed to be Gaussian:
\begin{equation}
P(w_{0},w_{a})\propto\exp(-c_{1}w_{0}^{2}-c_{2}w_{a}^{2}-c_{3}w_{0}w_{a}-c_{4}w_{0}-c_{5}w_{a})\,,
\label{eq:joint pdf}
\end{equation}
where the $c_{i}$ coefficients depend on the specific supernovae data set (Pantheon+, Union3 or DESY5) and can be calculated once we extract the equations for the constant-density contours presented in the latest DESI release \cite{DESI:2025zgx}. Consider now a concrete quintessence model (for instance, the quadratic hilltop potential with $k=1$). For every initial value of the field, $\varphi_{i}$, we obtained a pair of best-fit coordinates $(w_{0},w_{a})$. Therefore, the set of initial field values defined a curve $w_{a}(w_{0})$, which was approximately a line in most cases. Given this correspondence, the probability density function can be reduced to a single-variable function
\begin{equation}
p(w_{0})\propto P(w_{0},w_{a}(w_{0}))\,.
\label{eq:pdf single variable}
\end{equation}
(Since most curves $w_a(w_0)$ are linear, this suffices for our analysis. To be precise, there is also a correction factor of the form $\sqrt{1+w_a'(w_0)^2}$.)
Since there is typically a maximum value for $w_{0}$, $w_{0,\rm max}$, we will truncate $p(w_{0})$ at that point, meaning that $p(w_{0}>w_{0,\rm max})=0$. As described in \cite{Bayat:2025xfr}, even if $w_{0}=-1$ defines the left boundary of the domain, we extend the distribution to smaller values in order to assign a probability to the cosmological constant:
\begin{eqnarray}
p_{\rm null}&=&\frac{1}{\mathcal{N}}\int_{-\infty}^{-1}dw_{0}\,p(w_{0})\,,\\
\mathcal{N}&=&\int_{-\infty}^{w_{0,\rm max}}dw_{0}\,p(w_{0})\,.
\label{eq:cosmological constant probability}
\end{eqnarray}
Finally, one can translate this probability into a number of standard deviations as
\begin{equation}
N_{\sigma}=\sqrt{2}\erf^{-1}(1-2p_{\rm null})\,,
\label{eq:number standard deviations}
\end{equation}
which is the usual way of quantifying the statistical tension.

The results are given in the tables at the end of the paper.
We see that the number of standard deviations that the cosmological constant is in tension with the data is somewhat reduced compared to those reported by DESI in the entire $w_0,w_a$ plane.
An example is the cosine potential, which provides one of the better agreements between theory and observations. As can be seen in the table, for $k=3$, we have the number of standard deviations as
$N_\sigma\approx1.5,\,2.1,\,3.1$, for Pantheon-Plus, Union3, and DESY5 data sets, respectively. This is somewhat worse than the standard deviations reported by DESI in the entire $w_0,w_a$ plane of $N_\sigma=2.8,\,3.8,\,4.2$, for each data set, respectively. 

Although the tensions remain moderate in these models, we should also bear in mind that this involves new parameters, so there should be an additional penalty. Altogether, the tension of a cosmological constant in these models is arguably not large enough to significantly favor them over the null hypothesis.

Also, the fact that none of the models gives values in the $w_0,w_a$ plane that lies right through the data suggests that these simple quintessence constructions may not be quite right. This leads us to now go beyond the minimal coupling for a broader class of models.

\section{Non-Minimal Coupling}
\label{sec:nonminimal coupling}

In this section, we will consider scalar fields non-minimally coupled to gravity. The action is given by
\begin{equation}
S=\int d^{4}x\sqrt{-g}\left[\frac{M_{p}^{2}}{2}R-\frac{1}{2}g^{\mu\nu}\partial_{\mu}\varphi\partial_{\nu}\varphi-V(\varphi)-\frac{1}{2}\xi\varphi^{2}R+\mathcal{L}_{\rm m}\right]\,,
\label{eq:action non-minimal}
\end{equation}
where $\xi$ is the non-minimal coupling parameter. The energy-momentum tensor of the scalar field reads
\begin{equation}
T_{\mu\nu}=\partial_{\mu}\varphi\partial_{\nu}\varphi-g_{\mu\nu}\left[\frac{1}{2}g^{\alpha\beta}\partial_{\alpha}\varphi\partial_{\beta}\varphi+V(\varphi)\right]+\xi\left(G_{\mu\nu}+g_{\mu\nu}\Box-\nabla_{\mu}\nabla_{\nu}\right)\varphi^{2}\,,
\label{eq:Tmunu}
\end{equation}
where $G_{\mu\nu}$ is the Einstein tensor.
For an FLRW background,  
one finds the following expressions for the energy density and the pressure:
\begin{equation}
\rho_{\varphi}=\frac{1}{2}\dot{\varphi}^{2}+V(\varphi)+3H\xi\left(H\varphi^{2}+2\varphi\dot{\varphi}\right)\,,
\label{eq:energy density phi2}
\end{equation}
\begin{equation}
p_{\varphi}=\frac{1}{2}\dot{\varphi}^{2}-V(\varphi)
+\xi\left\{-2\dot{\varphi}^{2}+2\varphi V'(\varphi)+\left[H^{2}+2\left(\xi-\frac{1}{6}\right)R\right]\varphi^{2}+2H\varphi\dot{\varphi}\right\}\,.
\label{eq:pressure phi}
\end{equation}
Note that a non-vanishing coupling parameter allows, in principle, for NEC violation ($p_{\varphi}+\rho_{\varphi}<0$). Therefore, if the value of $\xi$, the scalar potential and the initial conditions are appropriate, the dark energy equation of state can cross the phantom divide $w_{\varphi}=-1$ from below and provide a good fit to the DESI data \cite{Ye:2024ywg,Wolf:2024stt}.

The evolution of the scalar field is found by numerically solving the equation of motion
\begin{equation}
\ddot{\varphi}+3H\dot{\varphi}+V'(\varphi)+\xi R\varphi=0\,,
\label{eq:eom non-minimal}
\end{equation}
with the Ricci scalar
\begin{equation}
R=\frac{\rho_{\rm m}+\rho_{\varphi}-3p_{\varphi}}{M_{p}^{2}}\,.
\label{eq:Ricci}
\end{equation}
Also, we have the Friedmann equation (\ref{eq:Friedmann}).

As in section \ref{sec:models}, the initial value of the scalar field, $\varphi_{i}$, is chosen freely, and the initial velocity, $\dot{\varphi}_{i}$, is zero. The initial matter energy density is chosen to be much higher than the initial dark energy density; we pick the value $10^6$ in our numerics. In order to do this, we first find the initial Hubble rate $H_{i}$ as a function of the initial matter density $\rho_{\rm m,i}$. Using Eqs. (\ref{eq:Friedmann}) and (\ref{eq:energy density phi2}), one can show that
\begin{equation}
H_{i}=\frac{\xi\varphi_{i}\dot{\varphi}_{i}+\sqrt{\xi^{2}\varphi_{i}^{2}\dot{\varphi}_{i}^{2}+\frac{1}{3}\left(M_p^2-\xi\varphi_{i}^{2}\right)\left(\frac{1}{2}\dot{\varphi}_{i}^{2}+V(\varphi_{i})+\rho_{\rm m,i}\right)}}{M_{p}^{2}-\xi\varphi_{i}^{2}}\,.
\label{eq:initial Hubble rate}
\end{equation}
Then we substitute this into (\ref{eq:energy density phi2}), set $\rho_{\varphi,i}=10^{-6}\rho_{\rm m,i}$ and solve for $\rho_{\rm m,i}$.
We note that in order for this form of dark energy to be very subdominant in the early universe can require significant fine-tuning of the initial conditions.

\subsection{Fifth Force Constraints}

The evolution of light, non-minimally coupled scalar fields is severely constrained by different tests of gravity. 

Firstly, the presence of a light scalar leads to a new fifth force in the Solar System. To derive this, let us expand around the homogeneous field $\varphi_0$ today as
\begin{equation}
    \varphi=\varphi_0+\delta\varphi\,.
\end{equation}
Then there is a linear term in the Lagrangian of the form
\begin{equation}
    \Delta\mathcal{L}=-\xi\varphi_0\,\delta\varphi\,R
    = \left(\xi\,\varphi_0\over M_p^2\right)\,\delta\varphi\,T\,,
\end{equation}
where in the second step we have used the trace of the Einstein equations for matter $R=-T/M_p^2$ (so this is only valid on the equations of motion). This linear term means that fluctuations in the scalar are coupled to matter via an effective coupling $g=\xi\varphi_0/M_p^2$.
This scalar $\delta\varphi$ mediates a fifth force between non-relativistic matter, but does not couple to light. In the weak field regime, we can express its consequences via the post-parameterized Newtonian parameter $\gamma$ which is defined via the weak field metric Lagrangian $ds^2=-(1+2\phi_N)dt^2+(1+2\gamma\phi_N)|d{\bf x}|^2$. This leads to
\begin{equation}
    \gamma={1-2g^2 M_p^2\over 1+2g^2M_p^2}\,.
\end{equation}
The leading bound on $\gamma$ comes from measurements of the Shapiro time delay of radio signals from the Cassini probe of $|\gamma-1|<2.3\times 10^{-5}$. Inserting the above value for $g$,
this leads to the following bound
\begin{equation}
\frac{|\xi\varphi_{0}|}{2.4\times 10^{-3}M_{p}}<1\,.
\label{eq:solar system constraint}
\end{equation}
We will refer to this bound as the 
``Solar System constraint''.
Defining
\begin{equation}
s(t)=\frac{\xi\varphi(t)}{2.4\times 10^{-3}M_{p}}\,,
\label{eq:s}
\end{equation}
the constraint reads $|s_{0}|<1$.

\subsection{Gravitational Coupling Constraints}

Secondly, as one can easily check from the action (\ref{eq:action non-minimal}), the strength of gravity is controlled by an effective gravitational constant which depends on time\footnote{In \cite{Rossi:2019lgt,Braglia:2020iik,Braglia:2020auw,FrancoAbellan:2023gec}, similar models with early time variation of the gravitational constant have been analyzed and tested against observations with a focus on their potential to alleviate the Hubble tension.}: 
\begin{equation}
G_{\rm eff}=\frac{G}{1-\xi\left(\frac{\varphi}{M_{p}}\right)^{2}}\,.
\label{eq:effective G}
\end{equation}
The time variation of this function is bounded \cite{Damour:1992kf,Damour:1993id,Uzan:2024ded}: 
\begin{equation}
\Bigg|\frac{\dot{G}_{\rm eff}}{G_{\rm eff}}\Bigg|_{0}\lesssim10^{-12}\text{\, yr}^{-1}\,.
\label{eq:gravitational constant constraint}
\end{equation}
Taking the time derivative of (\ref{eq:effective G}), we get
\begin{equation}
\Bigg|\frac{\dot{G}_{\rm eff}}{G_{\rm eff}}\Bigg|_{0}=\Bigg|\frac{2\xi\varphi\dot{\varphi}}{M_{p}^{2}-\xi\varphi^{2}}\Bigg|_{0}\lesssim10^{-12}\text{\, yr}^{-1}\,.
\label{eq:Gdot over G}
\end{equation}
We will refer to this bound as the ``effective gravitational constant constraint''.

The recent study \cite{Wolf:2025jed} suggests that the values of $\xi$ and $\varphi$ today ($\varphi_{0}$) that are required to explain the DESI data are inconsistent with these bounds, implying that the theory (\ref{eq:action non-minimal}) would have to be supplemented with extra ingredients. However, in what follows, we present an example (albeit fine-tuned) that seems to comply with the above constraints.

\subsection{Choice of Frame}

A subtlety is the choice of frame that we use to compare the predictions of our theory to the DESI data. As it stands, the action (\ref{eq:action non-minimal}) is written in the so-called Jordan frame. In this frame, the matter fields are minimally coupled to the Jordan metric. By means of a Weyl transformation of the metric and a field redefinition, one could rewrite the theory in the Einstein frame and get an action which looks like the standard action for a scalar field minimally coupled to gravity. However, the couplings of the matter fields to the new metric (the Einstein metric) are now $\varphi$-dependent; which manifests as a fifth force. 
Some discussion on the relation between the frames includes
Refs. \cite{Faraoni:2004pi,Faraoni:2010pgm}. 

An important related question is which of the frames is appropriate to directly test the theory against observations. While both frames are valid (since they are just related by field redefinitions), the Jordan frame seems to be more convenient for a direct comparison. The matter fields move on geodesics of the Jordan metric in the Jordan frame, but matter fields do not move on geodesics of the Einstein metric in the Einstein frame due to the additional fifth force. An implicit assumption behind the derivation of the usual distance-redshift and magnitude-redshift relations is that matter follows geodesics of the FLRW metric. The redshift is defined as $z=-1+\lambda_{0}/\lambda_{e}$, where $\lambda_{0}$ is the wavelength of the light from a distant source we observe today and $\lambda_{e}$ is supposed to be the wavelength of the light when it was emitted by the source. The value of $\lambda_{e}$ is actually determined from the light emitted by atoms in an experiment on Earth, and we assume this is the wavelength with which light was emitted by the distant source in the past. However, this interpretation would not be correct in general in the Einstein frame, because atomic transition frequencies in this frame are $\varphi$-dependent due to the non-minimal coupling of the matter fields to the metric. Therefore, in order to compare the predictions of the Einstein-frame action to observations, one would have to take this fact into account. 
Ref.~\cite{Chiba:2013mha} provides more details.



\subsection{Results}

\begin{figure}[t]
\centering
\includegraphics[scale=0.64]{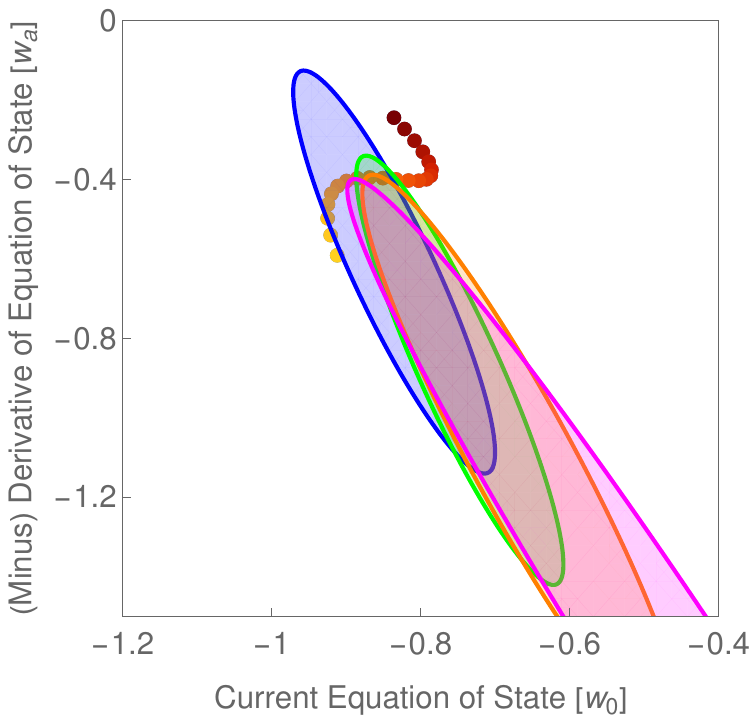}
\includegraphics[scale=0.61]{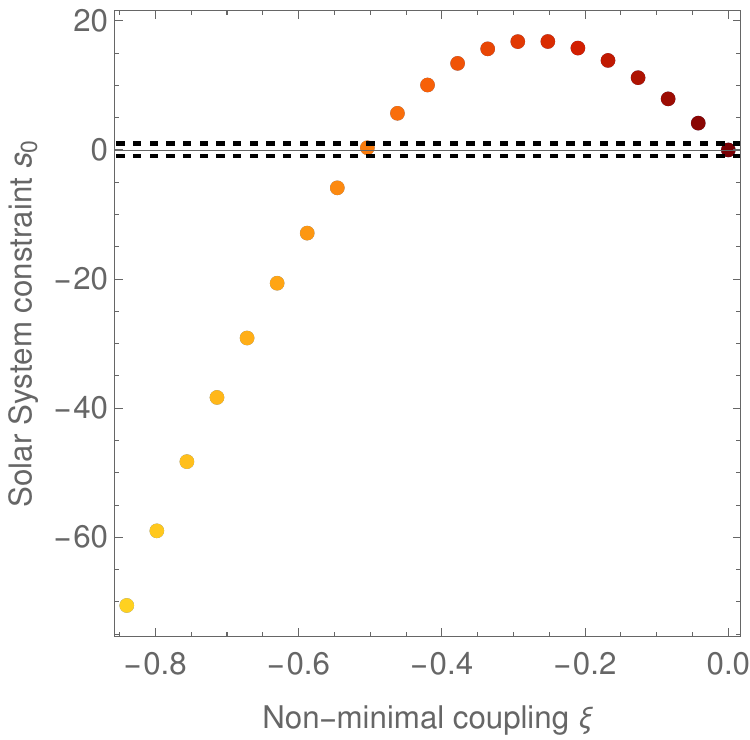}
\includegraphics[scale=0.66]{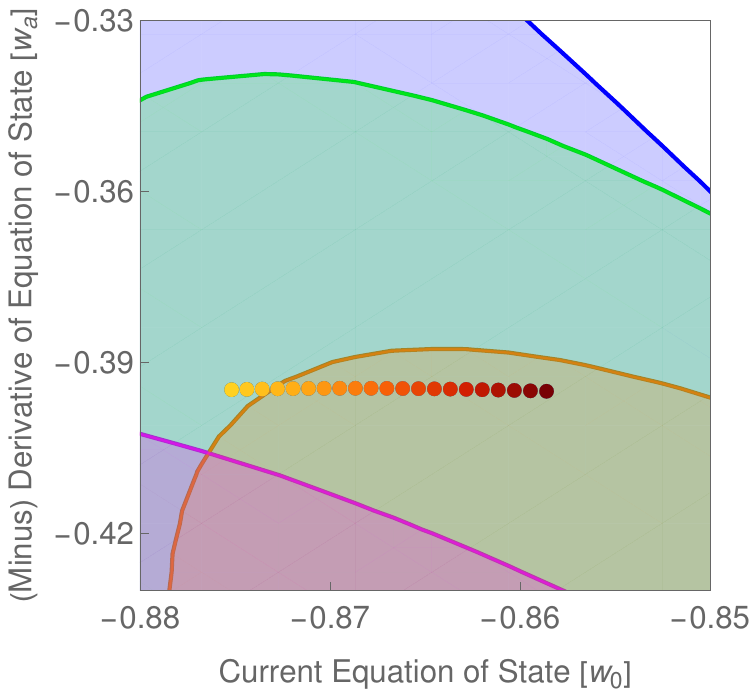}
\includegraphics[scale=0.61]{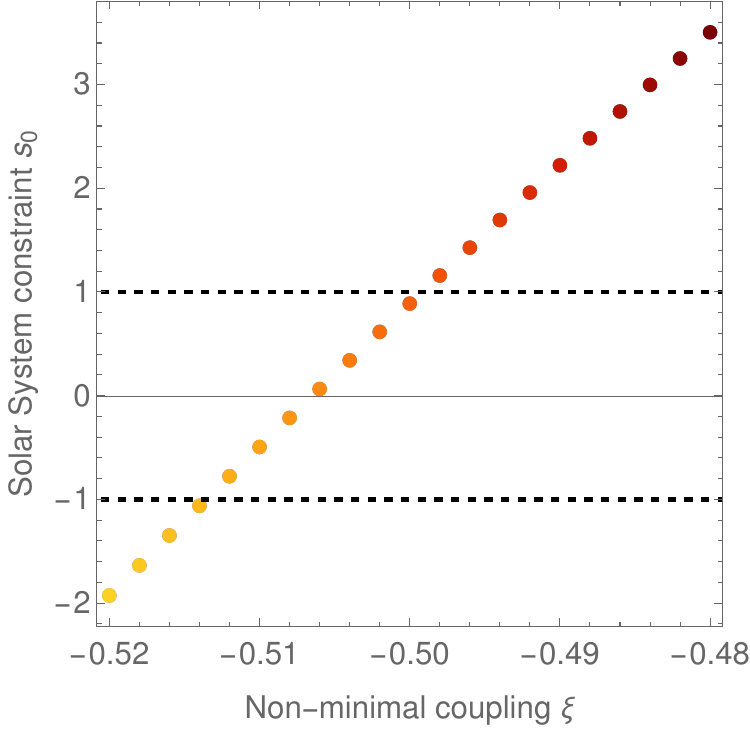}
\caption{Top left: DESI $2\sigma$ contours and ($w_0$,$w_a$) points for the values of non-minimal couplings $\xi$ indicated on the right. Top right: Solar System constraint (\ref{eq:solar system constraint}) for different values of $\xi$. The dashed lines indicate the lower and upper bounds (-1 and 1). 
Bottom panels are a zoomed in version focusing on $-0.52<\xi<-0.48$.
These plots are for $k=1,\,\varphi_i=0.08M_p$.}
\label{fig:nonmin}
\end{figure}

Here we present an example that satisfies the above fifth force restrictions. Moreover, its CPL parameters $w_{0}, w_{a}$ will lie inside of one or more of the $2\sigma$ DESI contours. We considered the simple linear potential 
\begin{equation}
V(\varphi)=V_{0}\left(1+\frac{k\,\varphi}{M_{p}}\right)\,.
\label{eq:tanh potential}
\end{equation}
In this model, $k$ acts as a parameter, in addition to the initial condition $\varphi_i$. However, to illustrate the basic idea, we will focus here on the following values:
\begin{equation}
k=1,\,\,\,\,\,\,\varphi_{i}=0.08 M_{p}\,.
\label{eq:initial phi}
\end{equation}
In Fig. \ref{fig:nonmin}, we display the CPL parameters in the $w_0$,$w_a$ plane for several values of the non-minimal coupling $\xi$. Also shown is the Solar System constraint (\ref{eq:solar system constraint}), which is only satisfied for a small range of $\xi$ values around 
$\xi\approx-0.5\,.$
A zoom in on this region is also shown in the figure.

\begin{figure}[t]
\centering
\includegraphics[scale=0.55]{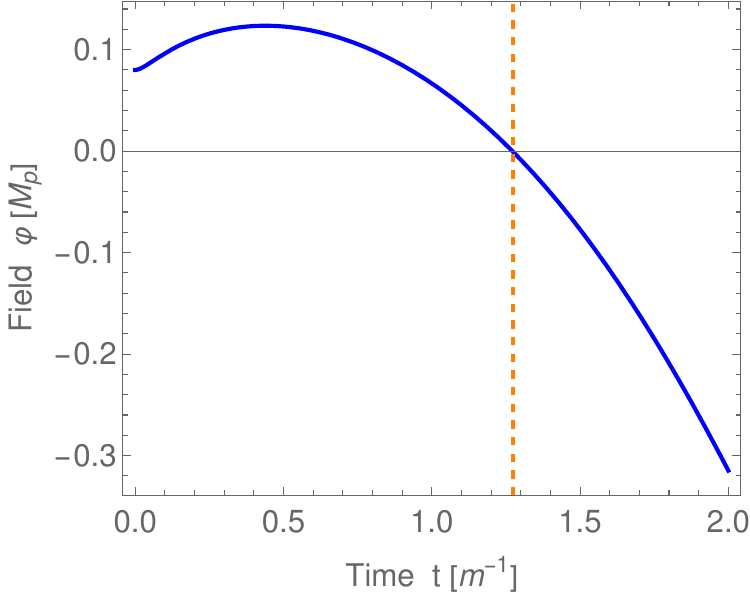}\,\,
\includegraphics[scale=0.54]{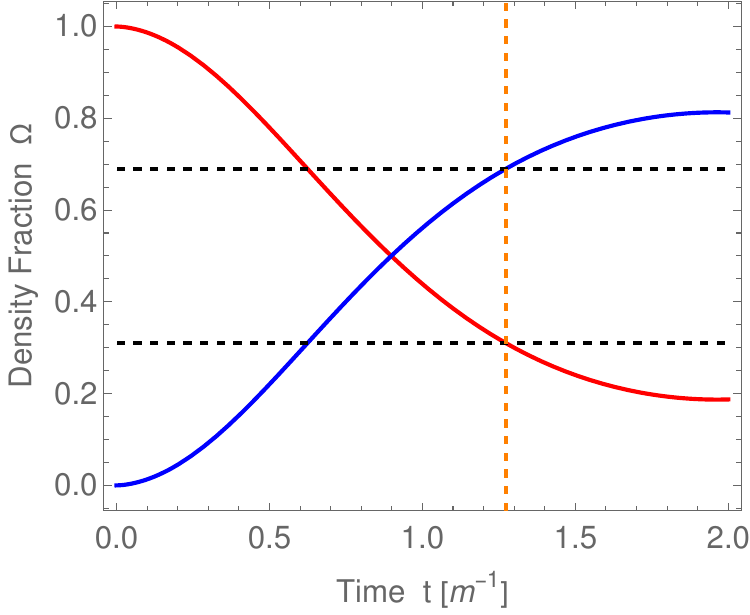}\\
\vspace{0.5cm}
\includegraphics[scale=0.61]{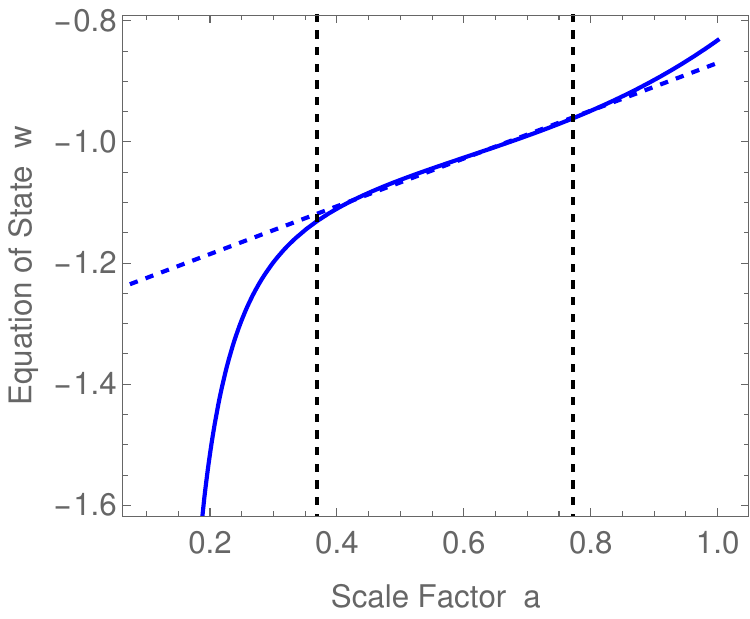}
\caption{
Top left: Value of the scalar field as a function of time. 
Top right: Fractional energy densities of matter (red) and dark energy (blue).
The vertical dashed line (orange) corresponds to the current time, defined as the moment at which the dark energy density fraction becomes $\Omega_{DE,0}=0.69$, which in this case is $t_0\approx 1.27 m^{-1}$. 
Bottom: Dark energy equation of state $w$ as a function of the scale factor (solid curve) and CPL fit (blue dashed line). The vertical lines indicate the range over which DESI is most sensitive. The present time is at $a=1$.
These plots are for the non-minimally coupled model with $k=1,\,\varphi_i=0.08M_p,\,\xi=-0.506$.}
\label{fig:nmc-tanh-field}
\end{figure}

\begin{figure}[h!]
\centering
\includegraphics[scale=0.6]{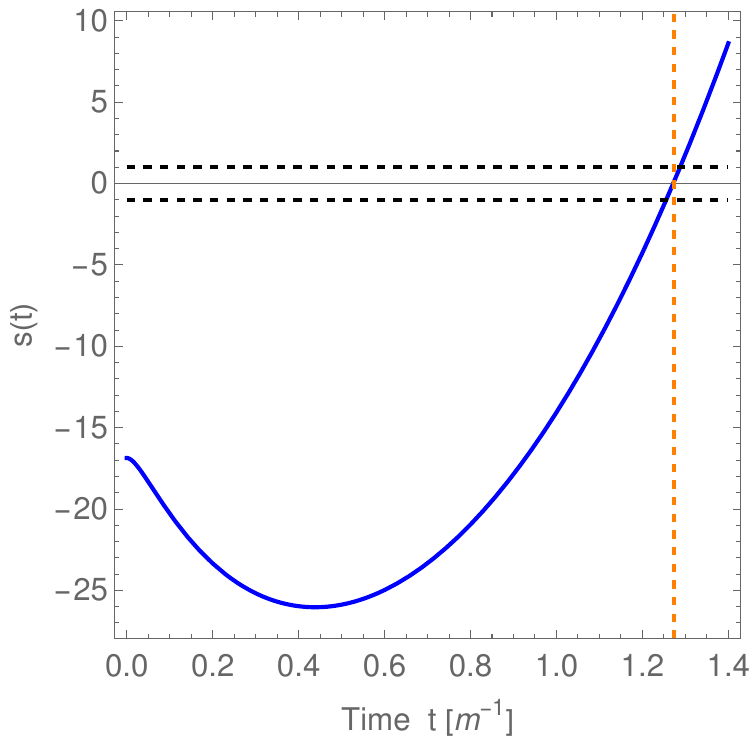}
\includegraphics[scale=0.67]{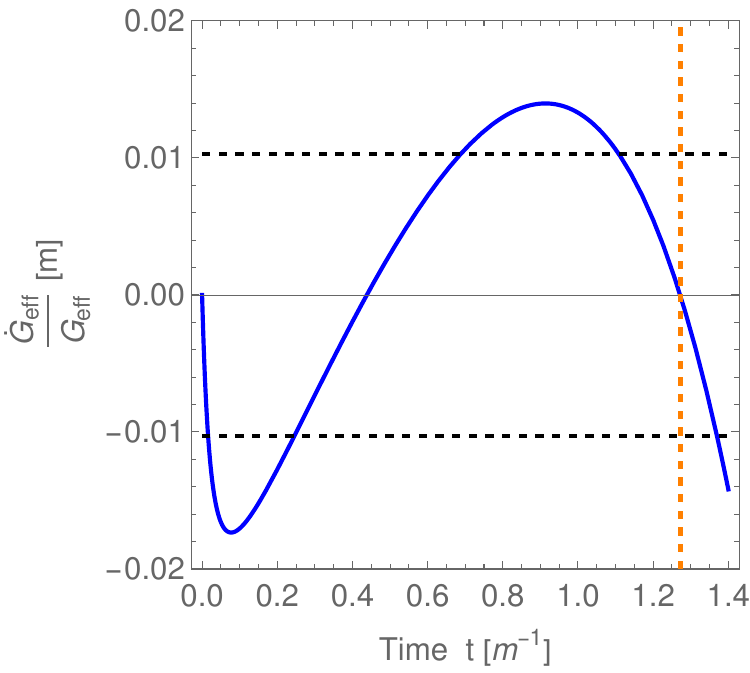}
\caption{Solar System (left) and effective gravitational constant (right) constraints versus time. The vertical dashed line (orange) corresponds to the current time $t_0\approx 1.27 m^{-1}$, and the horizontal dashed lines indicate the lower and upper bounds today.
This is for the non-minimally coupled model with $k=1,\,\varphi_i=0.08M_p,\,\xi=-0.506$.}
\label{fig:nmc-tanh-constraints}
\end{figure}

\begin{figure}[h]
\centering
\includegraphics[scale=0.62]{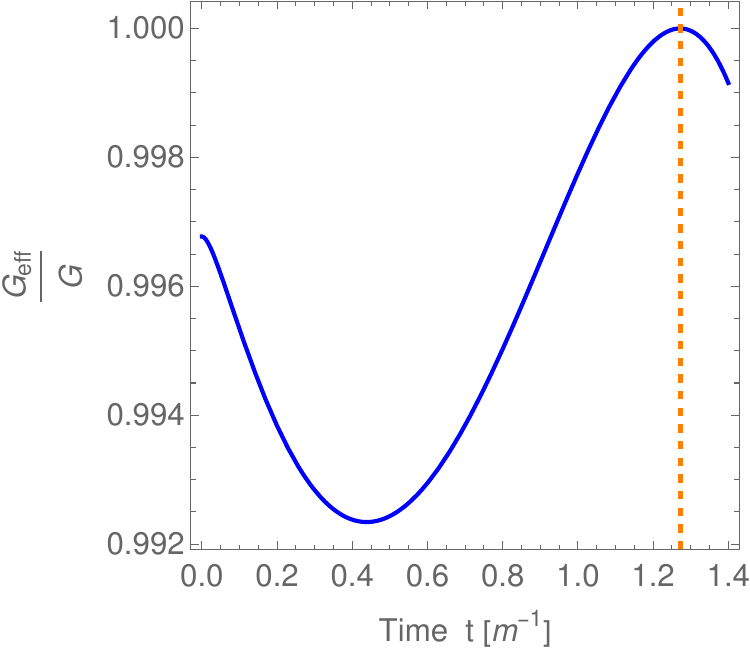}\,
\includegraphics[scale=0.64]{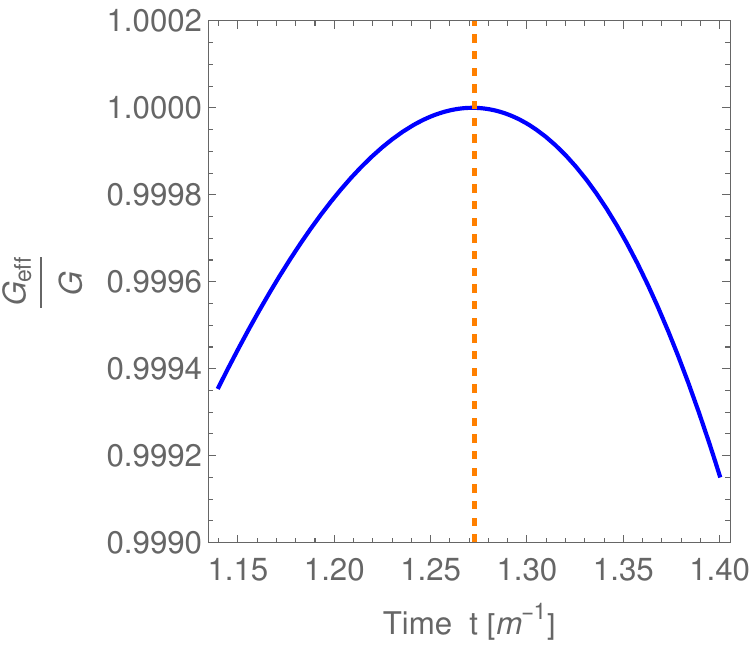}
\caption{Time evolution of the effective gravitational constant. The vertical dotted line corresponds to the current time $t_0\approx 1.27 m^{-1}$. The panel on the right is a zoomed-in view near the current time. 
}
\label{fig:Geff over G}
\end{figure}

The value of the field $\varphi$ as a function of time, as well as the matter and dark energy density fractions and the equation of state, are shown in Fig.~\ref{fig:nmc-tanh-field}
for the particular case $\xi=-0.506$. The constraints in this case are plotted in Fig. \ref{fig:nmc-tanh-constraints}. In order to directly test the gravitational constant constraint (\ref{eq:Gdot over G}), we must adapt it to our units. We are measuring time in units of $m^{-1}$; let us call this dimensionless time $\tilde{t}\equiv m\,t$. Solving the equations numerically, we can output a dimensionless Hubble parameter today as
\begin{equation}
\tilde{H}_0={1\over a}{da\over d\tilde{t}}\Bigg{|}_0
\end{equation}
where the physical Hubble parameter is $H_0=\tilde{H}_0\,m$. 
For the specific case $\xi=-0.506$, we find $\tilde{H}_0\approx 0.71$.
Using the measured value of Hubble parameter $H_{0}\approx6.91\times 10^{-11}\text{ yr}^{-1}$ \cite{Planck:2018jri},
we get $m\approx 9.73\times 10^{-11}\text{ yr}^{-1}$,
so the dimensionless right-hand-side of (\ref{eq:gravitational constant constraint}) and (\ref{eq:Gdot over G}) is $10^{-12}\text{ yr}^{-1}/m\approx 0.0103$. This is the bound indicated in the right panel of Fig. \ref{fig:nmc-tanh-constraints}. Finally, the time evolution of $G_{\rm eff}/G$ is shown in Fig. \ref{fig:Geff over G}.

We have carefully chosen parameters (fine-tuned) that satisfy the current solar system and time evolution bounds today.
For the solar system bound, this is because $\varphi$ is ``accidentally'' small today, though it was large in both the past and the future in these models, as seen in Fig.~\ref{fig:nmc-tanh-field}.
As far as we are aware, this is consistent with observations.
For the $\dot{G}_{\rm eff}/G_{\rm eff}$ bound, we note that it is slightly outside the current bound ($10^{-12}\,\mbox{yr}^{-1}$) in the past. Whether this is in tension with observations is not clear to us.

Finally, we  note that in our construction, nothing prevents the scalar field from dominating the energy density of the universe at redshifts  higher than the one corresponding to our initial condition. The shape of the potential and the coupling to the Ricci scalar chosen in this example need to be modified to some extent in order for dark energy to be subdominant at early times, including recombination, matter-radiation equality and earlier.



\clearpage

\section{Discussion}
\label{sec:conclusions}

In this work, we have explored a large range of quintessence models in light of the 2025 DESI data on dark energy. For minimally coupled scalar fields, we have found that the statistical significance is reduced compared to that reported by the DESI analysis in the entire $w_0,w_a$ plane (with equation of state of dynamical dark energy $w=w_0+(1-a)w_a$). While some models show reasonable agreement, none are able to reach inside the 2$\sigma$ region of the data. 

For non-minimally scalar fields, one can arrange for an improved fit to the data. However, this implies a new fifth force and a dynamical value for the effective Newton's gravitational coupling $G$. Both are tightly constrained by tests of gravity. For some carefully selected values, we can escape bounds from the Solar System by having the fifth force ``accidentally" be suppressed today. The bound on the evolution of Newton's coupling is able to be obeyed today, though there are moments in the history of the universe where our existing example is outside the bounds; see right panel in Fig.~\ref{fig:nmc-tanh-constraints}. Whether this is in tension with other bounds is less clear. From Fig.~\ref{fig:Geff over G} this implies a shifted value of $G$ in the very early universe, including Big Bang Nucleosynthesis. However, in this example, the shift is only at the level $\sim 0.3\%$ (we have not directly included radiation in our analysis to be certain, 
and as mentioned earlier, some modification of the theory is required to properly handle the very early universe). It would be interesting to provide further constraints on these kinds of model in the future.

Overall, generic models of quintessence can provide moderate improvement in fitting the DESI data compared to a cosmological constant. However, the presence of additional parameters means that there should be a penalty in such models. For very narrow ranges of non-minimally coupled models (fine-tuned) one can provide significant improvement in fitting the data, as well as evading some basic tests of gravity. Further work can involve a more systematic exploration of the full parameter space.

\section{Acknowledgments}

M.~P.-H. is supported in part by National Science 
Foundation grants PHY-2310572 and PHY-2419848.
D.~J.-A. is supported in part by National Science 
Foundation grant 
PHY-2419848.
We thank the VERSE program at Tufts for support. 
We thank David Kaiser for discussion.

\clearpage

\bibliography{DESI2025-bibliography}

\clearpage

\begin{table}
\begin{tabular}{|l|cc|cc|cc|cc|}
\hline
\multicolumn{9}{|c|}{\textbf{Model}: $V(\varphi)=V_{0}\left(1-\frac{k^{2}}{2M_{p}^2}\varphi^{2}\right)$} \\ 
\hline
& \multicolumn{2}{c|}{\textbf{Pan-Plus}}
& \multicolumn{2}{c|}{\textbf{Union3}}
& \multicolumn{2}{c|}{\textbf{DESY5}}
& \multicolumn{2}{c|}{\textbf{CMB}} \\
\cline{2-9}
& $p_{\text{null}}$ & $N_\sigma$ 
& $p_{\text{null}}$ & $N_\sigma$ 
& $p_{\text{null}}$ & $N_\sigma$ 
& $p_{\text{null}}$ & $N_\sigma$ \\
\hline
$k$=1 & 0.088 & 1.352 & 0.030 & 1.882 & 0.002 & 2.920 & 0.471 & 0.072 \\
$k$=1.2 & 0.086 & 1.366 & 0.029 & 1.902 & 0.002 & 2.936 & 0.464 & 0.089 \\
$k$=1.4 & 0.083 & 1.383 & 0.027 & 1.925 & 0.002 & 2.954 & 0.457 & 0.109 \\
$k$=1.6 & 0.081 & 1.400 & 0.026 & 1.949 & 0.001 & 2.973 & 0.448 & 0.130 \\
$k$=1.8 & 0.078 & 1.418 & 0.024 & 1.973 & 0.001 & 2.993 & 0.440 & 0.151 \\
$k$=2 & 0.075 & 1.437 & 0.023 & 1.994 & 0.001 & 3.013 & 0.433 & 0.169 \\
$k$=2.5 & 0.069 & 1.483 & 0.022 & 2.005 & 0.001 & 3.049 & 0.424 & 0.191 \\
$k$=3 & 0.066 & 1.510 & 0.029 & 1.893 & 0.001 & 2.977 & 0.443 & 0.144\\
$k$=4 & 0.099 & 1.287 & 0.089 & 1.350 & 0.009 & 2.366 & 0.560 & - \\
$k$=5 & 0.250 & 0.675 & 0.274 & 0.600 & 0.076 & 1.429 & 0.729 & - \\
$k$=6 & 0.453 & 0.118 & 0.493 & 0.017 & 0.243 & 0.697 & 0.841 & - \\
$k$=7 & 0.672 & - & 0.708 & - & 0.500 & 0.001 & 0.894 & - \\
$k$=8 & 0.819 & - & 0.841 & - & 0.715 & - & 0.926 & - \\
$k$=9 & 0.640 & - & 0.668 & - & 0.636 & - & 0.689 & - \\
$k$=10 & 0.871 & - & 0.884 & - & 0.881 & - & 0.885 & - \\
\hline
\end{tabular}
\end{table}

\begin{table}
\begin{tabular}{|l|cc|cc|cc|cc|}
\hline
\multicolumn{9}{|c|}{\textbf{Model}: $V(\varphi)=V_{0}\left(1-\frac{k^{4}}{4M_{p}^4}\varphi^{4}\right)$} \\ 
\hline
& \multicolumn{2}{c|}{\textbf{Pan-Plus}} 
& \multicolumn{2}{c|}{\textbf{Union3}}
& \multicolumn{2}{c|}{\textbf{DESY5}} 
& \multicolumn{2}{c|}{\textbf{CMB}} \\
\cline{2-9}
& $p_{\text{null}}$ & $N_\sigma$ 
& $p_{\text{null}}$ & $N_\sigma$ 
& $p_{\text{null}}$ & $N_\sigma$ 
& $p_{\text{null}}$ & $N_\sigma$ \\
$k$=0.5 & 0.082 & 1.390 & 0.027 & 1.920 & 0.002 & 2.952 & 0.457 & 0.107 \\
$k$=1 & 0.076 & 1.433 & 0.024 & 1.972 & 0.001 & 2.998 & 0.439 & 0.153 \\
$k$=1.6 & 0.071 & 1.469 & 0.028 & 1.913 & 0.001 & 2.987 & 0.447 & 0.133 \\
$k$=2 & 0.072 & 1.464 & 0.036 & 1.794 & 0.002 & 2.884 & 0.470 & 0.074 \\
$k$=4 & 0.181 & 0.911 & 0.191 & 0.875 & 0.037 & 1.782 & 0.672 & - \\
$k$=8 & 0.602 & - & 0.637 & - & 0.401 & 0.251 & 0.897 & - \\
\hline
\end{tabular}
\end{table}

\begin{table}
\begin{tabular}{|l|cc|cc|cc|cc|}
\hline
\multicolumn{9}{|c|}{\textbf{Model}: $V(\varphi)=V_{0}\left(1-\frac{k^{2}}{2M_{p}^2}\varphi^{2} +\frac{\lambda k^4}{16M_{p}^4}\varphi^{4}\right)$} \\ 
\hline
& \multicolumn{2}{c|}{\textbf{Pan-Plus}} 
& \multicolumn{2}{c|}{\textbf{Union3}}
& \multicolumn{2}{c|}{\textbf{DESY5}} 
& \multicolumn{2}{c|}{\textbf{CMB}} \\
\cline{2-9}
& $p_{\text{null}}$ & $N_\sigma$ 
& $p_{\text{null}}$ & $N_\sigma$ 
& $p_{\text{null}}$ & $N_\sigma$ 
& $p_{\text{null}}$ & $N_\sigma$ \\
$k$=1,\,\,\,\, $\lambda$ = 0.32 & 0.089 & 1.348 & 0.030 & 1.875 & 0.002 & 2.914 & 0.474 & 0.065 \\
$k$=1.5, $\lambda$ = 0.32 & 0.083 & 1.388 & 0.027 & 1.930 & 0.002 & 2.958 & 0.455 & 0.114 \\
$k$=2,\,\,\,\, $\lambda$ = 0.32 & 0.076 & 1.434 & 0.023 & 1.993 & 0.001 & 3.009 & 0.433 & 0.169 \\
$k$=1,\,\,\,\, $\lambda$ = 0.48 & 0.089 & 1.347 & 0.031 & 1.871 & 0.002 & 2.911 & 0.475 & 0.062 \\
$k$=1.5, $\lambda$ = 0.48  & 0.083 & 1.387 & 0.027 & 1.927 & 0.002 & 2.956 & 0.456 & 0.111 \\
$k$=2,\,\,\,\, $\lambda$ = 0.48  & 0.076 & 1.433 & 0.023 & 1.991 & 0.001 & 3.006 & 0.433 & 0.168 \\
$k$=1,\,\,\,\, $\lambda$ = 0.64 & 0.089 & 1.345 & 0.031 & 1.867 & 0.002 & 2.908 & 0.477 & 0.058 \\
$k$=1.5, $\lambda$ = 0.64 & 0.083 & 1.385 & 0.027 & 1.923 & 0.002 & 2.953 & 0.457 & 0.108 \\
$k$=2,\,\,\,\, $\lambda$ = 0.64 & 0.076 & 1.432 & 0.023 & 1.988 & 0.001 & 3.004 & 0.434 & 0.166 \\
$k$=1,\,\,\,\, $\lambda$ = 0.8 & 0.090 & 1.343 & 0.031 & 1.862 & 0.002 & 2.904 & 0.478 & 0.054 \\
$k$=1.5, $\lambda$ = 0.8 & 0.083 & 1.383 & 0.028 & 1.919 & 0.002 & 2.950 & 0.458 & 0.104 \\
$k$=2,\,\,\,\, $\lambda$ = 0.8 & 0.076 & 1.430 & 0.024 & 1.985 & 0.001 & 3.002 & 0.435 & 0.164 \\
$k$=1,\,\,\,\, $\lambda$ = 0.96 & 0.090 & 1.340 & 0.032 & 1.857 & 0.002 & 2.900 & 0.480 & 0.050 \\
$k$=1.5, $\lambda$ = 0.96 & 0.084 & 1.382 & 0.028 & 1.914 & 0.002 & 2.947 & 0.460 & 0.101 \\
$k$=2,\,\,\,\, $\lambda$ = 0.96 & 0.077 & 1.429 & 0.024 & 1.981 & 0.001 & 2.999 & 0.436 & 0.162 \\
\hline
\end{tabular}
\end{table}

\begin{table}
\begin{tabular}{|l|cc|cc|cc|cc|}
\hline
\multicolumn{9}{|c|}{\textbf{Model}: $V(\varphi)=V_{0}\left(1-\frac{k^{2}}{2M_{p}^2}\varphi^{2} +\frac{k^4}{16M_{p}^4}\varphi^{4}\right)$} \\ 
\hline
&\multicolumn{2}{c|}{\textbf{Pan-Plus}} 
& \multicolumn{2}{c|}{\textbf{Union3}}
& \multicolumn{2}{c|}{\textbf{DESY5}} 
& \multicolumn{2}{c|}{\textbf{CMB}} \\
\cline{2-9}
& $p_{\text{null}}$ & $N_\sigma$ 
& $p_{\text{null}}$ & $N_\sigma$ 
& $p_{\text{null}}$ & $N_\sigma$ 
& $p_{\text{null}}$ & $N_\sigma$ \\
\hline
$k$=0.5 & 0.095 & 1.310 & 0.035 & 1.813 & 0.002 & 2.865 & 0.494 & 0.015 \\
$k$=1 & 0.090 & 1.340 & 0.032 & 1.855 & 0.002 & 2.899 & 0.480 & 0.049 \\
$k$=1.2 & 0.088 & 1.355 & 0.030 & 1.877 & 0.002 & 2.917 & 0.473 & 0.068 \\
$k$=1.4 & 0.085 & 1.372 & 0.029 & 1.901 & 0.002 & 2.936 & 0.465 & 0.089 \\
$k$=1.5 & 0.084 & 1.381 & 0.028 & 1.913 & 0.002 & 2.946 & 0.460 & 0.100 \\
$k$=1.6 & 0.082 & 1.390 & 0.027 & 1.926 & 0.002 & 2.956 & 0.456 & 0.111 \\
$k$=1.8 & 0.079 & 1.409 & 0.025 & 1.953 & 0.001 & 2.977 & 0.446 & 0.136 \\
$k$=2 & 0.077 & 1.428 & 0.024 & 1.981 & 0.001 & 2.999 & 0.436 & 0.161 \\
$k$=2.5 & 0.070 & 1.476 & 0.020 & 2.047 & 0.001 & 3.052 & 0.412 & 0.222 \\
$k$=3 & 0.064 & 1.520 & 0.020 & 2.056 & 0.001 & 3.089 & 0.406 & 0.239 \\
$k$=3.5 & 0.061 & 1.546 & 0.026 & 1.942 & 0.001 & 3.021 & 0.426 & 0.187 \\
$k$=4 & 0.071 & 1.467 & 0.049 & 1.651 & 0.003 & 2.720 & 0.489 & 0.029 \\
$k$=4.5 & 0.102 & 1.270 & 0.094 & 1.316 & 0.010 & 2.321 & 0.565 & - \\
$k$=5 & 0.172 & 0.946 & 0.183 & 0.904 & 0.035 & 1.810 & 0.660 & - \\
$k$=5.5 & 0.260 & 0.645 & 0.286 & 0.566 & 0.083 & 1.384 & 0.735 & - \\
$k$=6 & 0.259 & 0.648 & 0.285 & 0.567 & 0.083 & 1.385 & 0.734 & - \\
$k$=6.5 & 0.511 & - & 0.552 & - & 0.304 & 0.514 & 0.863 & - \\
$k$=7 & 0.617 & - & 0.655 & - & 0.427 & 0.183 & 0.892 & - \\
$k$=7.5 & 0.715 & - & 0.746 & - & 0.555 & - & 0.928 & - \\
$k$=8 & 0.786 & - & 0.811 & - & 0.657 & - & 0.946 & - \\
$k$=8.5 & 0.847 & - & 0.858 & - & 0.748 & - & 0.946 & - \\
$k$=9 & 0.893 & - & 0.903 & - & 0.820 & - & 0.966 & - \\
$k$=9.5 & 0.927 & - & 0.935 & - & 0.877 & - & 0.978 & - \\
$k$=10 & 0.950 & - & 0.957 & - & 0.916 & - & 0.984 & - \\
\hline
\end{tabular}
\end{table}

\begin{table}
\newpage
\begin{tabular}{|l|cc|cc|cc|cc|}
\hline
\multicolumn{9}{|c|}{\textbf{Model}: $V(\varphi)=\frac{V_{0}}{2}\left(1+\cos\left(\frac{\sqrt{2}\,k\,\varphi}{M_{p}}\right)\right)$} \\ 
\hline
& \multicolumn{2}{c|}{\textbf{Pan-Plus}} 
& \multicolumn{2}{c|}{\textbf{Union3}}
& \multicolumn{2}{c|}{\textbf{DESY5}} 
& \multicolumn{2}{c|}{\textbf{CMB}} \\
\cline{2-9}
& $p_{\text{null}}$ & $N_\sigma$ 
& $p_{\text{null}}$ & $N_\sigma$ 
& $p_{\text{null}}$ & $N_\sigma$ 
& $p_{\text{null}}$ & $N_\sigma$ \\
\hline
$k$=0.5 & 0.096 & 1.307 & 0.035 & 1.806 & 0.002 & 2.859 & 0.497 & 0.007 \\
$k$=1 & 0.091 & 1.336 & 0.032 & 1.846 & 0.002 & 2.892 & 0.483 & 0.042 \\
$k$=1.2 & 0.088 & 1.351 & 0.031 & 1.868 & 0.002 & 2.910 & 0.476 & 0.060 \\
$k$=1.4 & 0.086 & 1.368 & 0.029 & 1.892 & 0.002 & 2.929 & 0.467 & 0.082 \\
$k$=1.6 & 0.083 & 1.387 & 0.028 & 1.918 & 0.002 & 2.950 & 0.458 & 0.105 \\
$k$=1.8 & 0.080 & 1.406 & 0.026 & 1.946 & 0.001 & 2.972 & 0.448 & 0.130 \\
$k$=2 & 0.077 & 1.426 & 0.024 & 1.974 & 0.001 & 2.994 & 0.438 & 0.155 \\
$k$=2.5 & 0.070 & 1.474 & 0.020 & 2.044 & 0.001 & 3.048 & 0.413 & 0.221 \\
$k$=3 & 0.064 & 1.519 & 0.017 & 2.109 & 0.001 & 3.097 & 0.388 & 0.286 \\
$k$=3.5 & 0.060 & 1.552 & 0.023 & 1.995 & 0.001 & 3.063 & 0.414 & 0.218 \\
$k$=4 & 0.067 & 1.498 & 0.043 & 1.719 & 0.003 & 2.795 & 0.473 & 0.068 \\
$k$=4.5 & 0.084 & 1.376 & 0.070 & 1.473 & 0.006 & 2.509 & 0.528 & - \\
$k$=5 & 0.134 & 1.107 & 0.136 & 1.096 & 0.020 & 2.049 & 0.616 & - \\
$k$=5.5 & 0.216 & 0.784 & 0.236 & 0.718 & 0.058 & 1.575 & 0.702 & - \\
$k$=6 & 0.348 & 0.391 & 0.383 & 0.298 & 0.148 & 1.047 & 0.789 & - \\
$k$=6.5 & 0.474 & 0.066 & 0.514 & - & 0.264 & 0.631 & 0.848 & - \\
$k$=7 & 0.599 & - & 0.638 & - & 0.406 & 0.238 & 0.894 & - \\
$k$=7.5 & 0.624 & - & 0.662 & - & 0.437 & 0.159 & 0.902 & - \\
$k$=8 & 0.716 & - & 0.748 & - & 0.558 & - & 0.930 & - \\
$k$=8.5 & 0.829 & - & 0.850 & - & 0.722 & - & 0.960 & - \\
$k$=9 & 0.879 & - & 0.895 & - & 0.800 & - & 0.969 & - \\
$k$=9.5 & 0.915 & - & 0.925 & - & 0.857 & - & 0.977 & - \\
$k$=10 & 0.936 & - & 0.945 & - & 0.894 & - & 0.978 & - \\
\hline
\end{tabular}
\end{table}

\clearpage

\begin{table}
\begin{tabular}{|l|cc|cc|cc|cc|}
\hline
\multicolumn{9}{|c|}{\textbf{Model}: $V(\varphi)=V_0\left(\varphi\over M_p\right)$} \\ 
\hline
& \multicolumn{2}{c|}{\textbf{Pan-Plus}} 
& \multicolumn{2}{c|}{\textbf{Union3}} 
& \multicolumn{2}{c|}{\textbf{DESY5}} 
& \multicolumn{2}{c|}{\textbf{CMB}} \\
\cline{2-9}
& $p_{\text{null}}$ & $N_\sigma$ 
& $p_{\text{null}}$ & $N_\sigma$ 
& $p_{\text{null}}$ & $N_\sigma$ 
& $p_{\text{null}}$ & $N_\sigma$ \\
& 0.096 & 1.307 & 0.035 & 1.816 & 0.002 & 2.867 & 0.494 & 0.014 \\
\hline
\end{tabular}
\end{table}

\begin{table}
\begin{tabular}{|l|cc|cc|cc|cc|}
\hline
\multicolumn{9}{|c|}{\textbf{Model}: $V(\varphi)=\frac{V_0}{2}\left(\varphi\over M_p\right)^2$} \\ 
\hline
& \multicolumn{2}{c|}{\textbf{Pan-Plus}} 
& \multicolumn{2}{c|}{\textbf{Union3}}
& \multicolumn{2}{c|}{\textbf{DESY5}}
& \multicolumn{2}{c|}{\textbf{CMB}} \\
\cline{2-9}
& $p_{\text{null}}$ & $N_\sigma$ 
& $p_{\text{null}}$ & $N_\sigma$ 
& $p_{\text{null}}$ & $N_\sigma$ 
& $p_{\text{null}}$ & $N_\sigma$ \\
& 0.097 & 1.296 & 0.037 & 1.791 & 0.002 & 2.847 & 0.502 & - \\
\hline
\end{tabular}
\end{table}

\begin{table}
\begin{tabular}{|l|cc|cc|cc|cc|}
\hline
\multicolumn{9}{|c|}{\textbf{Model}: $V(\varphi)=\frac{V_0}{4}\left(\varphi\over M_p\right)^4$} \\ 
\hline
& \multicolumn{2}{c|}{\textbf{Pan-Plus}}
& \multicolumn{2}{c|}{\textbf{Union3}}
& \multicolumn{2}{c|}{\textbf{DESY5}}
& \multicolumn{2}{c|}{\textbf{CMB}} \\
\cline{2-9}
& $p_{\text{null}}$ & $N_\sigma$ 
& $p_{\text{null}}$ & $N_\sigma$ 
& $p_{\text{null}}$ & $N_\sigma$ 
& $p_{\text{null}}$ & $N_\sigma$ \\
& 0.098 & 1.291 &  0.038 & 1.778 &  0.002 & 2.837 & 0.506 & - \\
\hline
\end{tabular}
\end{table}

\begin{table}
\begin{tabular}{|l|cc|cc|cc|cc|}
\hline
\multicolumn{9}{|c|}{\textbf{Model}: $V(\varphi)=V_{0}\,\exp\!\left(-{\frac{k^2\varphi^2}{2M_{p}^2}}\right)$} \\ 
\hline
& \multicolumn{2}{c|}{\textbf{Pan-Plus}}
& \multicolumn{2}{c|}{\textbf{Union3}} 
& \multicolumn{2}{c|}{\textbf{DESY5}} 
& \multicolumn{2}{c|}{\textbf{CMB}} \\
\cline{2-9}
& $p_{\text{null}}$ & $N_\sigma$ 
& $p_{\text{null}}$ & $N_\sigma$ 
& $p_{\text{null}}$ & $N_\sigma$ 
& $p_{\text{null}}$ & $N_\sigma$ \\
$k$=1 & 0.092 & 1.326 & 0.034 & 1.823 & 0.002 & 2.875 & 0.490 & 0.025 \\
$k$=1.2 & 0.090 & 1.342 & 0.032 & 1.847 & 0.002 & 2.894 & 0.482 & 0.044 \\
$k$=1.4 & 0.087 & 1.360 & 0.031 & 1.873 & 0.002 & 2.915 & 0.473 & 0.067 \\
$k$=1.5 & 0.085 & 1.370 & 0.030 & 1.886 & 0.002 & 2.925 & 0.469 & 0.078 \\
$k$=1.6 & 0.084 & 1.379 & 0.029 & 1.900 & 0.002 & 2.937 & 0.464 & 0.091 \\
$k$=1.8 & 0.081 & 1.399 & 0.027 & 1.929 & 0.002 & 2.959 & 0.454 & 0.116 \\
$k$=2 & 0.078 & 1.419 & 0.025 & 1.959 & 0.001 & 2.983 & 0.443 & 0.143 \\
$k$=2.5 & 0.071 & 1.469 & 0.021 & 2.031 & 0.001 & 3.039 & 0.416 & 0.211 \\
$k$=3 & 0.065 & 1.515 & 0.018 & 2.099 & 0.001 & 3.090 & 0.391 & 0.276 \\
$k$=3.5 & 0.060 & 1.555 & 0.017 & 2.120 & 0.001 & 3.130 & 0.383 & 0.298 \\
$k$=4 & 0.059 & 1.560 & 0.028 & 1.904 & 0.001 & 2.985 & 0.430 & 0.177 \\
$k$=4.5 & 0.067 & 1.496 & 0.046 & 1.687 & 0.003 & 2.756 & 0.476 & 0.059 \\
$k$=5 & 0.103 & 1.263 & 0.097 & 1.299 & 0.011 & 2.298 & 0.567 & - \\
$k$=5.5 & 0.174 & 0.940 & 0.186 & 0.894 & 0.036 & 1.795 & 0.661 & - \\
$k$=6 & 0.263 & 0.633 & 0.290 & 0.552 & 0.086 & 1.366 & 0.736 & - \\
$k$=6.5 & 0.400 & 0.252 & 0.439 & 0.153 & 0.194 & 0.865 & 0.817 & - \\
$k$=7 & 0.537 & - & 0.577 & - & 0.332 & 0.433 & 0.869 & - \\
$k$=7.5 & 0.627 & - & 0.666 & - & 0.442 & 0.146 & 0.884 & - \\
$k$=8 & 0.730 & - & 0.759 & - & 0.577 & - & 0.925 & - \\
$k$=8.5 & 0.801 & - & 0.826 & - & 0.681 & - & 0.942 & - \\
$k$=9 & 0.863 & - & 0.881 & - & 0.781 & - & 0.947 & - \\
$k$=9.5 & 0.894 & - & 0.898 & - & 0.823 & - & 0.956 & - \\
$k$=10 & 0.928 & - & 0.937 & - & 0.891 & - & 0.963 & - \\
\hline
\end{tabular}
\end{table}

\begin{table}
\begin{tabular}{|l|cc|cc|cc|cc|}
\hline
\multicolumn{9}{|c|}{\textbf{Model}: $V(\varphi)=V_0\left(1+\frac{k^2\varphi^2}{2M_{p}^2}\right)^{\!-1}$} \\ 
\hline
& \multicolumn{2}{c|}{\textbf{Pan-Plus}} 
& \multicolumn{2}{c|}{\textbf{Union3}}
& \multicolumn{2}{c|}{\textbf{DESY5}} 
& \multicolumn{2}{c|}{\textbf{CMB}} \\
\cline{2-9}
& $p_{\text{null}}$ & $N_\sigma$ 
& $p_{\text{null}}$ & $N_\sigma$ 
& $p_{\text{null}}$ & $N_\sigma$ 
& $p_{\text{null}}$ & $N_\sigma$ \\
$k$=1 & 0.128 & 1.136 & 0.104 & 1.258 & 0.011 & 2.302 & 0.608 & - \\
$k$=2 & 0.081 & 1.399 & 0.028 & 1.904 & 0.002 & 2.943 & 0.459 & 0.102 \\
$k$=3 & 0.066 & 1.503 & 0.019 & 2.067 & 0.001 & 3.068 & 0.401 & 0.250 \\
$k$=4 & 0.057 & 1.580 & 0.015 & 2.183 & 0.001 & 3.156 & 0.356 & 0.369 \\
$k$=5 & 0.052 & 1.628 & 0.016 & 2.147 & 0.001 & 3.178 & 0.363 & 0.350 \\
$k$=6 & 0.095 & 1.310 & 0.088 & 1.355 & 0.009 & 2.360 & 0.543 & - \\
$k$=7 & 0.300 & 0.525 & 0.327 & 0.449 & 0.111 & 1.219 & 0.718 & - \\
$k$=8 & 0.533 & - & 0.567 & - & 0.382 & 0.300 & 0.710 & - \\
$k$=9 & 0.720 & - & 0.745 & - & 0.645 & - & 0.809 & - \\
$k$=10 & 0.778 & - & 0.761 & - & 0.673 & - & 0.855 & - \\
\hline
\end{tabular}
\end{table}

\begin{table}
\begin{tabular}{|l|cc|cc|cc|cc|}
\hline
\multicolumn{9}{|c|}{\textbf{Model}: $V(\varphi)=V_0{\left(1+\frac{k^2\varphi^2}{M_{p}^2}\right)}^{\!-1/2}$} \\ 
\hline
& \multicolumn{2}{c|}{\textbf{Pan-Plus}} 
& \multicolumn{2}{c|}{\textbf{Union3}}
& \multicolumn{2}{c|}{\textbf{DESY5}} 
& \multicolumn{2}{c|}{\textbf{CMB}} \\
\cline{2-9}
& $p_{\text{null}}$ & $N_\sigma$ 
& $p_{\text{null}}$ & $N_\sigma$ 
& $p_{\text{null}}$ & $N_\sigma$ 
& $p_{\text{null}}$ & $N_\sigma$ \\
$k$=2 & 0.085 & 1.371 & 0.038 & 1.778 & 0.002 & 2.861 & 0.491 & 0.023 \\
$k$=3 & 0.068 & 1.490 & 0.021 & 2.029 & 0.001 & 3.043 & 0.413 & 0.219 \\
$k$=4 & 0.058 & 1.571 & 0.015 & 2.160 & 0.001 & 3.141 & 0.364 & 0.348 \\
$k$=5 & 0.052 & 1.623 & 0.013 & 2.240 & 0.001 & 3.199 & 0.331 & 0.437 \\
$k$=6 & 0.049 & 1.656 & 0.011 & 2.288 & 0.001 & 3.235 & 0.309 & 0.499 \\
$k$=7 & 0.047 & 1.676 & 0.010 & 2.318 & 0.001 & 3.256 & 0.291 & 0.551 \\
$k$=8 & 0.046 & 1.689 & 0.010 & 2.339 & 0.001 & 3.270 & 0.269 & 0.615 \\
$k$=9 & 0.045 & 1.697 & 0.009 & 2.351 & 0.001 & 3.279 & 0.266 & 0.625 \\
\hline
\end{tabular}
\end{table}

\end{document}